\documentclass[preprint]{aastex}
\usepackage{appendix}
\usepackage{ulem}
\usepackage{multicol}
\usepackage{xcolor}

\shorttitle{Type Ibn Supernova ASASSN-14ms} \shortauthors{Wang et al.}

\def\gsim{\;\lower4pt\hbox{${\buildrel\displaystyle >\over\sim}$}\;} \def\lsim{\;\lower4pt\hbox{${\buildrel\displaystyle <\over\sim}$}\;} \def\grls{\;\lower4pt\hbox{${\buildrel\displaystyle >\over <}$}\;}

\begin{document}

\title{ASASSN-14ms: the Most Energetic Known Explosion of a Type Ibn Supernova and its Physical Origin}

\author{Xiaofeng Wang\altaffilmark{1,2}, Weili Lin\altaffilmark{1}, Jujia Zhang\altaffilmark{3,4}, Tianmeng Zhang\altaffilmark{5,6}, Yongzhi Cai\altaffilmark{1}, Kaicheng Zhang\altaffilmark{1}, Alexei V. Filippenko\altaffilmark{7,8}, Melissa Graham\altaffilmark{9}, Keiichi Maeda\altaffilmark{10}, Jun Mo\altaffilmark{1}, Danfeng Xiang\altaffilmark{1}, Gaobo Xi\altaffilmark{1}, Shengyu Yan\altaffilmark{1}, Lifan Wang\altaffilmark{11}, Lingjun Wang\altaffilmark{12}, Koji Kawabata\altaffilmark{13}, and Qian Zhai\altaffilmark{3,4}}

\altaffiltext{1}{Physics Department and Tsinghua Center for Astrophysics (THCA), Tsinghua University, Beijing 100084, China; wang\_xf@mail.tsinghua.edu.cn,linwl@mail.tsinghua.edu.cn}
\altaffiltext{2}{Beijing Planetarium, Beijing Academy of Science and Technology, Beijing 100044, China}
\altaffiltext{3}{Yunnan Observatories, Chinese Academy of Sciences, Kunming 650216, China}
\altaffiltext{4}{Key Laboratory for the Structure and Evolution of Celestial Objects, Chinese Academy of Sciences, Kunming 650216, China}
\altaffiltext{5}{Key Laboratory of Optical Astronomy, National Astronomical Observatories, Chinese Academy of Sciences, Beijing 100101, China}
\altaffiltext{6}{School of Astronomy and Space Science, University of Chinese Academy of Sciences, Beijing 101408, China}
\altaffiltext{7}{Department of Astronomy, University of California, Berkeley, CA 94720-3411, USA}
\altaffiltext{8}{Miller Institute for Basic Research in Science, University of California, Berkeley, CA 94720, USA}
\altaffiltext{9}{Department of Astronomy, University of Washington, Box 351580,  Seattle, WA 98195-1580, USA}
\altaffiltext{10}{Department of Astronomy, Kyoto University, Kitashirakawa-Oiwake-cho, Sakyo-ku, Kyoto 606-8502, Japan}
\altaffiltext{11}{Physics and Astronomy Department, Texas A\&M University, College Station, TX 77843, USA}
\altaffiltext{12}{Astroparticle Physics, Institute of High Energy Physics, Chinese Academy of Sciences, Beijing 100049, China}
\altaffiltext{13}{Department of Physical Science, Hirosima University, Kagamiyama, Higashi-Hiroshima, Hiroshima 739-8526, Japan}

\begin{abstract}
ASASSN-14ms may represent the most luminous Type Ibn supernova (SN~Ibn) ever detected, with an absolute $U$-band magnitude brighter than $-$22.0 mag and a total bolometric luminosity $> 1.0 \times 10^{44}$ erg s$^{-1}$ near maximum light. The early-time spectra of this SN are characterized by a blue continuum on which are superimposed narrow P~Cygni profile lines of He\,\textsc{i}, suggesting the presence of slowly moving ($\sim 1000$ km s$^{-1}$), He-rich circumstellar material (CSM). At 1--2 months after maximum brightness, the He\,\textsc{i} line profiles become only slightly broader, with blueshifted velocities of 2000--3000 km s$^{-1}$, consistent with the CSM shell being continuously accelerated by the SN light and ejecta. Like most SNe~Ibn, the light curves of ASASSN-14ms show rapid post-peak evolution, dropping by $\sim 7$ mag in the $V$ band over three months. Such a rapid post-peak decline and high luminosity can be explained with interaction between SN ejecta and helium-rich CSM of $0.9~M_{\odot}$ at a distance of $\sim 10^{15}$ cm. The CSM around ASASSN-14ms is estimated to originate from a pre-explosion event with a mass-loss rate of $6.7~M_\odot$~yr$^{-1}$ (assuming a velocity of $\sim 1000$ km s$^{-1}$), which is consistent with abundant He-rich material violently ejected during the late Wolf-Rayet (WN9-11 or Opfe) stage. After examining the light curves for a sample of SNe~Ibn, we find that the more luminous ones tend to have slower post-peak decline rates, reflecting that the observed differences may arise primarily from discrepancies in the CSM distribution around the massive progenitors.

\end{abstract}

\section{Introduction}
Type Ibn supernovae (SNe~Ibn) constitute a subtype of core-collapse events that are characterized by relatively narrow emission lines of He\,\textsc{i} in their spectra, suggesting the presence of helium-rich circumstellar material (CSM; \citealp{2000AJ....119.2303M}). They are rare and make up only about 1\% of core-collapse SNe (Pastorello et al. 2008a). The progenitors of SNe~Ibn are thought to be Wolf-Rayet (WR) stars that lost all of their hydrogen envelope and part of the helium shell prior to exploding. A prototype of such events is SN 2006jc which was observed to experience a pre-SN outburst \citep{2007ApJ...657L.105F, 2007Natur.447..829P}.

Over the past few years, $\sim 40$ SNe~Ibn were identified, and some have very good follow-up observations (see \citealp{2016MNRAS.456..853P, 2017ApJ...836..158H}, and references therein). Based on their early-time spectra, SNe~Ibn can be equally classified into the P~Cygni and emission subclasses \citep{2017ApJ...836..158H}, with the former being characterized by narrow P~Cygni profiles and the latter being dominated by broader emission lines. It is currently unclear whether these distinctions trace two different CSM configurations or a continuum of CSM properties. Moreover, inspection of their birthplace environments indicates that most SNe~Ibn tend to occur in star-forming regions of their host galaxies \citep{2015MNRAS.449.1954P, 2015A&A...580A.131T}, suggesting that they are associated with massive stars.

SNe~Ibn are found to be quite luminous among SNe; most of them have absolute peak magnitudes ranging roughly from $-18.5$ to $-20.0$ mag in the $R$ band. Their light curves show a large variety in pre- and post-maximum-brightness evolution (e.g., see Fig.~3 of \citealp{2016MNRAS.456..853P}). Their rise time can span a period from $\lesssim 1$ week to 2--3 weeks. Those with rapidly-rising light curves might be related to the mysterious, so-called ``fast evolving luminous transients'' (FELTs; e.g., Rest et al. 2018; Xiang et al. 2021). Some SNe~Ibn are found to show multiple light-curve peaks (e.g., \citealp{2014MNRAS.443..671G}) or a flattening in the post-peak light curve \citep{2015MNRAS.449.1941P}, while most of them tend to have very rapidly-declining light curves after maximum light (e.g., $\sim 8.0$ mag within the first 100 days; see also Hosseinzadeh et al. 2017). The observed diversity in the light curves indicates that the energy sources responsible for SNe~Ibn are complicated, possibly including radioactive decay, interaction of SN ejecta with CSM, magnetar, or some other unknown mechanisms.

ASASSN-14ms provides another rare opportunity to study the diverse properties of SNe~Ibn. This object was discovered on 2014 December 26.61 (UT dates are used throughout this paper) by the All Sky Automated Survey for SuperNovae (ASAS-SN; \citealp{2014ApJ...788...48S}) in SDSS J130408.52+521846.4, at 16.5 mag in the $V$ band (Kiyota et al. 2014). Our earliest spectrum of this object, obtained with the Lijiang 2.4~m telescope (+YFOSC) on 2014 December 29.84, showed a very blue continuum with weak, narrow P~Cygni He\,\textsc{i} lines, suggesting the Type Ibn classification. Its J2000 coordinates are $\alpha = 13^{\rm h}04^{\rm m}08.69^{\rm s}$ and $\delta = +52\degr 18 \arcmin 46\farcs 5$, approximately 1.1$\arcsec$ south and 0.8$\arcsec$ east from the center of SDSS J130408.52+521846.4 (see Fig.1). The host is very faint, with a $g$-band magnitude of $\sim 21.6$, and there is no previously published measurement for its recession velocity.

Owing to the existing small number of well-studied SNe~Ibn, we started follow-up observations of ASASSN-14ms immediately after obtaining the first classification spectrum. Although this SN was investigated by Vallely et al. (2018), we present more extensive observations and an independent analysis of it. In \S 2, we present our photometric and spectroscopic observations. The light curves and spectra are analyzed in \S 3, modeling of the light curve is given in \S 4, and we conclude in \S 5.

\section{Observations}
\subsection{Supernova ASASSN14ms}
We commenced optical photometry of ASASSN-14ms with different instruments, including the 0.8~m Tsinghua University-NAOC Telescope (hereafter TNT) at Xinglong Observatory in China \citep{2008ApJ...675..626W, 2012RAA....12.1585H}, the 2.4~m telescope of Yunnan Observatories at Lijiang Station (hereafter LJT; Wang et al. 2019) in China, and the 2.3~m Bok telescope (hereafter Bok; Williams et al. 2004) of NAOA in the USA. These observations spanned from about $-2$ days to +130 days from maximum brightness, and all of the data were reduced with standard IRAF\footnote{IRAF, the Image Reduction and Analysis Facility, is distributed by the National Optical Astronomy Observatory, which is operated by the Association of Universities for Research in Astronomy (AURA), Inc. under cooperative agreement with the National Science Foundation (NSF).} routines. The SN instrumental magnitudes were converted to the Johnsons $UBV$ \citep{1966CoLPL...4...99J} and Kron-Cousins $RI$ \citep{1981SAAOC...6....4C} systems, using flux calibrations of six field stars from the Sloan Digital Sky Survey (SDSS) Data Release 9 catalog as listed in Table 1. Table 2 lists the final flux-calibrated $UBVRI$ photometry of ASASSN-14ms. A few reports for upper limits and detection magnitudes in the $V$ band, obtained at earlier phases from the ASAS-SN and amateur astronomers, are also included.

In addition, ultraviolet (UV) and optical photometry from the Ultra-Violet/Optical Telescope (UVOT; \citealp{2005SSRv..120...95R}) onboard the {\it Neil Gehrels Swift Observatory} ($Swift$; \citealp{2004ApJ...611.1005G}) were available at two epochs around maximum light. The UVOT observations were obtained in the $uvw1$, $uvm2$, $uvw2$, $u$, $b$, and $v$ filters, with the images being reduced following that of the Swift Optical Ultraviolet Supernova Archive\cite[SOUSA;][]{2014Ap&SS.354...89B}. An aperture of 3 arcsec is applied to measure the SN flux, with the corresponding aperture correction based on an average point-spread function. The final UVOT magnitudes are listed in Table 3.

Spectroscopic observations were collected using the Lijiang 2.4~m telescope and YFOSC system \citep{2015RAA....15..918F}, the Xinglong 2.16~m telescope (+BFOSC), and the Keck 10~m telescope (+LRIS; Oke et al. 1995). All of the spectra were reduced using routine tasks within IRAF, and the flux was calibrated with spectrophotometric standard stars observed on the same nights. Telluric lines were removed from the spectra through comparison with the standard-star observations. A journal of observations is given in Table 4.

\subsection{Host Galaxy SDSS J130408.52+521846.4}
As redshift is not available for the galaxy SDSS J130408.52+521846.4 from the public catalogue, we collected the host-galaxy spectra from two telescopes. As the Keck spectrum was taken $\sim 50$ days after maximum brightness when the SN light was comparable to that of the host galaxy, it was also used to extract the host-galaxy spectrum to determine the redshift. Another spectrum of the host galaxy SDSS J130408.52+521846.4 was obtained on 2016 June 21.4 (t$\sim$ 6 months after the peak) with the 8.2~m Subaru telescope to cross-check the result from the Keck telescope. Figure 2 shows the host-galaxy spectra of ASASSN-14ms from Keck (blue) and Subaru (red). One can see that the narrow emission line at 6915~\AA, with a measured full width at half-maximum intensity (FWHM) of $\sim 300$ km s$^{-1}$, is clearly present in both spectra. Assuming this feature to be H$\alpha$, a redshift of $z = 0.0535$ can be inferred for SDSS J130408.52+521846.4. This value is consistent with $z = 0.0539$ determined from another narrow emission feature (at 3928~\AA) which can be attributed to [O\,\textsc{ii}] $\lambda$3727, as seen in the Keck spectrum (upper panel of Fig.2). We thus adopt a mean value of $z = 0.0537 \pm 0.0002$ for ASASSN-14ms. This redshift is consistent with that derived by \citet{2018MNRAS.475.2344V} from a late-time LBT MODS spectrum of SDSS J130408.52+521846.4 (obtained at t$\sim$ 876 days after maximum light). Assuming H$_{0}$ = 73 km s$^{-1}$ Mpc$^{-1}$, $\Omega_{M} = 0.27$, and $\Omega_{\Lambda} = 0.73$ \citep{2007ApJS..170..377S}, a distance modulus of $36.72 \pm 0.14$ mag can be obtained for ASASSN-14ms, which is adopted in the following analysis.

From the H$_{\alpha}$ luminosity of flux-calibrated Subaru spectrum, we estimated the star formation rate using the conversion relation from Kennicutt (1998), which is half of the value derived by \citet{2018MNRAS.475.2344V}, i.e.,$\sim$0.005 M$_{\odot}$ yr$^{-1}$. This difference is likely due to that the Subaru spectrum has relatively low signal-to-noise ratio. Modeling of the photometric spectral energy distribution for galaxy SDSS J130408.52+521846.4 reveals that this galaxy has a stellar mass $M_{*}$=2.6$^{+0.8}$$_{-1.5}$$\times$10$^8$ M$_{\odot}$ \citep{2018MNRAS.475.2344V}. This indicates a specific star formation rate of log(sSFR)$\sim-$10.11 M$_{\odot}$ yr$^{-1}$ for the host galaxy of ASASSN14ms well within the range of core collapse SNe (see Fig.9 of Xiang et al. 2021).

\section{Photometric and Spectroscopic Analysis}
\subsection{Light Curves}
Figure~3 shows the optical light curves of ASASSN-14ms, with two UV data points from {\it Swift}/UVOT overplotted. The earliest prediscovery image was taken on 2014 December 15.51, with a nondetection limit of $\sim 17.1$ mag in the $V$ band. The absence of early-phase data does not allow us to place stringent constraints on the explosion date and hence the rise time. The $V-$band light curve is better sampled around maximum light compared to other bands. Applying a polynomial fit to the $V$ light curve yields $V_{\rm max} = 16.42 \pm 0.04$ mag on JD 2,457,023.85. Likewise, the $B$-band light curve reaches a peak of $16.30 \pm 0.05$ mag at a similar phase, while the $U$-band light curve seems to have a slightly earlier peak with $U_{\rm max} \approx 14.9$ mag. Maximum brightness is not clear in the $R$ and $I$ bands because of sparse sampling. On the other hand, we notice that all of the light curves seem to experience three-stage post-peak evolution, with breaks occurring at $t \approx 20$ days and $t \approx 40$ days after maximum. During the period from $t \approx 20$ days to $t \approx 40$ days after maximum, the light curves declined at a relatively faster pace.

Taking the distance modulus $\mu = 36.72 \pm 0.14$ mag and a Galactic reddening of $E(B-V) = 0.01$ mag, we derive an absolute $V$-band peak magnitude of $-20.33 \pm 0.15$ mag for ASASSN-14ms. The corresponding values in $B$ and $U$ are $-20.46 \pm 0.15$ mag and $\gtrsim -21.9$ mag, respectively. With the available {\it Swift} $UV$ points, one can find that this object is also very luminous in the UV, with the luminosity in the $uvw1$, $uvm2$, and $uvw2$ bands being brighter than $-22.1$ mag. Detailed photometric parameters are listed in Table 3. These estimates indicate that ASASSN-14ms is one of the most luminous SNe~Ibn ever recorded (e.g., \citealp{2016MNRAS.456..853P, 2017ApJ...836..158H}), even without considering possible extinction due to the host galaxy.

Figure~4 shows the $V$-band light curve of ASASSN-14ms compared with those of well-observed SNe~Ibn such as SN 2006jc \citep{2007Natur.447..829P}, SN 1991D \citep{2002MNRAS.336...91B}, SN 2010al \citep{2015MNRAS.449.1921P}, SN 2014av \citep{2016MNRAS.456..853P}, 2015U\citep{2016MNRAS.461.3057S}, ASASSN-15ed \citep{2015MNRAS.453.3649P}, and 2019uo \citep{2020ApJ...889..170G}. The light curves of other subclasses are also plotted for comparison, including broad-lined Type Ic SN 1998bw \citep{1998Natur.395..670G}, Type Ia SN 2005cf \citep{2009ApJ...697..380W}, Type Ic SN 2007gr \citep{2014ApJ...790..120C}, and Type Ib SN 2009jf \citep{2011MNRAS.413.2583S}. It is readily seen that ASASSN-14ms and other comparison SNe~Ibn are characterized by fast post-peak declines. For example, the decline within the first 40 days is about 3.6 mag for ASASSN-14ms, 4.1 mag for ASASSN-15ed, 4.2 mag for SN 2010al, and 3.4 mag for SN 2014av.\footnote{The flattening behavior seen in SN 2014av after 20 days from maximum light is likely due to the interaction of the SN ejecta with the CSM from the progenitor.} In contrast, the corresponding magnitude decline is about 2.0, 2.1, 2.0, and 1.6 for SN 1998bw, SN 2005cf, SN 2007gr, and SN 2009jf (respectively). This large discrepancy in the post-maximum evolution indicates that SNe~Ibn are not primarily powered by radioactive decay in a relatively long phase after peak brightness (see discussions in \S 4) and their high luminosities are likely caused by other mechanisms.

Owing to the high luminosity of ASASSN-14ms, we further explore its possible connection with superluminous supernovae (SLSNe). To better quantify the discrepancy in light-curve evolution between ASASSN-14ms, other well-observed SNe~Ibn, and Type I/II SLSNe, we examine the absolute peak magnitude in $V$ ($M_{V,\mathrm{p}}$) and the magnitude decline measured within the first 30 days after the peak ($\Delta m_{30}(V)$). As shown in Figure~5, the absolute $V$ magnitude of most SNe~Ibn is in the approximate range $-17$ to $-20$ around maximum light. Among them, ASASSN-14ms is at the most luminous end with $M_{V,\mathrm{p}} \approx -20.5$ mag, close to the faint end of SLSNe. For SNe~Ibn and SLSNe, the decline rate measured within 30 days after the peak tends to show a correlation with the corresponding absolute peak magnitudes, with brighter objects having slower decline rates. Applying a linear fit to the observed data for SNe~Ibn and SLSNe yields $M_{V,\mathrm{p}} = -22.75^{+0.06}_{-0.02}+1.08^{+0.01}_{-0.02} \times \Delta m_{30}$. This trend suggests that these two subclasses of SNe may have a similar energy source. If their early-time radiation is dominated by ejecta-CSM interaction in both cases (see discussions in \S 4), the difference in decline rate might suggest that SNe~Ibn tend to have less ejecta mass and/or less extended CSM than the SLSNe. However, the establishment of this anti-correlation among SLSNe and SNe Ibn requires more statistically significant evidence.

In Figure~6, we compare the color evolution of ASASSN-14ms with that of the comparison SNe~Ibn. In the first 20 days, ASASSN-14ms seems to exhibit $B - V$ color evolution that is quite similar to that of SN 2010al and SN 2014av. After $t \approx 20$ days, both SN 2010al and SN 2014av tend to maintain a constant $B - V$ color of $\sim 0.6$ mag, whereas ASASSN-14ms evolves progressively redward until $t \approx 30$ days, subsequently evolving blueward and reaching a similar color as the former two events at $t \approx 50$ days. ASASSN-15ed may have a similar evolutionary trend as ASASSN-14ms, except that it is systematically redder by $\sim 0.2$ mag after $t \approx 10$ days. Among the comparison sample, ASASSN-15ed and SN 2006jc exhibit the fastest and slowest color evolution, respectively. One can see that SN 2010al and SN 2014av also show similar $R - I$ color evolution, with a reddening rate of $\lesssim 0.01$ mag d$^-1$ during the first 40 days. Slow $R - I$ color evolution is also found in ASASSN-14ms, but overall it is significantly bluer than SN 2010al, SN 2014av, and SN 2006jc. The $R-I$ evolution of SN 2015U might be consistent with that of ASASSN-14ms, but the comparison is limited by the large uncertainty in the former's color at $t \approx 10$--20 days.

\subsection{Spectroscopy}
Twelve optical spectra were obtained of ASASSN-14ms, covering the approximate phases $-2.5$ days to +50.0 days from maximum light; they are displayed in Figure~7. The earlier spectra are characterized by a very blue continuum, suggesting a high temperature for the emitting region at early phases and consistent with the strong UV emission revealed by {\it Swift} photometry obtained around the time of maximum light. Later, some narrow lines with P~Cygni profiles emerge in the spectra. The feature at $\sim 5900$~\AA, persisting in almost all of the spectra, is likely due to He\,\textsc{i} $\lambda$5876. Other helium lines can be also detected at 4470~\AA, 6700~\AA, and 7060~\AA (see dashed lines in Fig.~\ref{fig-specomp}). Note that these line profiles of helium (both absorption and emission components) get only slightly stronger from the near-maximum-light phase to $t \approx 50$ days after maximum brightness, indicating that they are primarily formed in a slowly-moving helium shell instead of in SN ejecta.

To further examine the spectral features of ASASSN-14ms, we compare it with different subtypes of core-collapse SNe in which the hydrogen envelope was stripped away before the explosion, as shown in Figure~8. The comparison objects include SNe~Ibn such as SN 2010al \citep{2015MNRAS.449.1921P} and ASASSN-15ed \citep{2015MNRAS.453.3649P}, the peculiar Type Ib SN 1991D \citep{2002MNRAS.336...91B}, and the normal Type Ib SN 2009jf \citep{2011MNRAS.413.2583S}. At one week after maximum light, ASASSN-14ms shows narrow P~Cygni profiles of helium lines at $\sim 4470$~\AA, 5860~\AA, and 6700~\AA\, similar to those seen in SN 2010al and ASASSN-15ed. In particular, all are found to show the ``M''-shaped feature at 6600--6700~\AA, which is likely formed by He\,\textsc{i} or Ne\,\textsc{i} (see the SYNOW fit shown in Figure~9). Note that the weak emission feature on the left side of He\,\textsc{i} $\lambda$6678 could be H$\alpha$, as similarly seen in SN 2010al and SN 2009jf, and it is not unexpected that some hydrogen mixed into the helium shell.  At about four weeks after peak brightness, the line profiles of the two comparison SNe~Ibn (SN 2010al and ASASSN-15ed) and the peculiar SN~Ib (SN 1991D) become apparently stronger (see the lower panel of Fig.~8), with the absorption component of He\,\textsc{i} $\lambda$5876 having a velocity of 5000--6500 km s$^{-1}$. In contrast, the corresponding blueshifted velocity is only $\sim 2000$ km s$^{-1}$ in ASASSN-14ms, suggesting that its He\,\textsc{i} line is still primarily formed from the CSM.

Figure~10 shows the temporal evolution of the line profiles of He\,\textsc{i} $\lambda$5876, He\,\textsc{i} $\lambda$6678, and O\,\textsc{i} $\lambda$7774. One can see that the absorption components of these lines tend to have blueshifted velocities increasing from $\sim 800$ km s$^{-1}$ at around maximum light to $\sim2000$ km s$^{-1}$ about 1 month later. Such an evolution in the above line profiles is similarly seen in a few spectra from \citet{2018MNRAS.475.2344V} (see gray spectra in Fig.~10). This indicates that the CSM was abundant with helium and oxygen, and that the CSM surrounding ASASSN-14ms was only slightly accelerated by the SN ejecta, suggesting that the CSM was either more distant or the mass was relatively larger.

To further constrain the elements in the ejecta of ASASSN-14ms, we use SYNAPPS and the companion code SYN++ (Thomas et al. 2011) to fit the $t \approx +24.5$~d spectrum. We first fit the spectrum automatically with the SYNAPPS through the reduced chi-square technique, which involves more than a dozen of ions. We then ignored some unimportant ions and used 6 main ions (including He I, C I, O I, Ne I, Ca II and Fe II) to fit the main features in this spectrum, with the best-fit result (judged by eye) shown in Figure~9 and the parameters listed in Table 5. From the best-fit result, one can see that the intermediate-mass elements such as He\,\textsc{i}, O\,\textsc{i}, Ne\,\textsc{i}, and Ca\,\textsc{ii} (and perhaps C\,\textsc{i}) can be identified in the spectrum of ASASSN-14ms. Note that the Ne I features around 6000 \AA\ can also be seen in Type Ib supernova SN 1991D (Bennett et al. 2002). Abundant Fe\,\textsc{ii} elements are also needed to reproduce the prominent absorptions around 4500~\AA\ and 5100~\AA. However, it is not clear which elements lead to the broad trough near 3600~\AA.

Around maximum light, the expansion velocity inferred from the absorption minima of He\,\textsc{i} lines for ASASSN-14ms is measured to be $\sim 1000$ km s$^{-1}$. As shown in Figure~11, other SNe~Ibn show similarly low velocities, as measured at comparable phases. In comparison, normal SNe~Ib have expansion velocities ranging from $\sim 7000$ km s$^{-1}$ to $\sim 12,000$ km s$^{-1}$, much larger than the typical velocity derived for SNe~Ibn. Moreover, these two subclasses with prominent He\,\textsc{i} features have distinct peak luminosities, with SNe~Ibn being on average much more luminous than SNe~Ib. We note that SN 1991D shows the P~Cygni line profiles lying between those of normal SNe~Ib and SNe~Ibn, with a blueshifted velocity of $\sim 5000$ km s$^{-1}$; it may be regarded as a transitional object linking these two subclasses. This suggests that SNe~Ib and SNe~Ibn may not be distinctly different, and their observed differences are likely caused by the surrounding CSM.

\section{Modeling the Light Curve}

We constructed the Galactic extinction-corrected spectral energy distributions (SEDs) of ASASSN-14ms using the multiband data, and then fit them with blackbody curves to obtain the bolometric luminosity $L_\mathrm{bol}$ and the effective temperature $T_\mathrm{eff}$\footnote{For accuracy, only the SEDs constructed from data in at least five filters, and obtained by fitting the $UBVRI$ light curves with quintic polynomial functions, are used to estimate the bolometric luminosity and the effective temperature. Those SEDs observed after MJD 57080 cannot provide a stringent constraint on the effective temperature, and they are thus not included in our analysis}. We estimated the effective temperature before MJD 57022 by extrapolating linearly from the temperature evolution yielded from the UVOT observations, and obtained the early-time bolometric luminosity based on the inferred effective temperature and the $V/g$-band magnitudes reported in our paper and \citet{2018MNRAS.475.2344V}. As shown in the upper panel of Figure~12, our bolometric light curve shows a smooth and monotonic decay, unlike that reported by \citet{2018MNRAS.475.2344V} who proposed a possible detection of a secondary feature near MJD 57050. We suspect that this discrepancy is likely due to the large uncertainty associated with their photometry. From the declining light curve, we infer that ASASSN-14ms had a peak luminosity higher than $10^{44}$~erg~s$^{-1}$, brighter than that of ordinary SNe~Ibn \citep{2015MNRAS.449.1921P, 2015MNRAS.449.1941P, 2015MNRAS.449.1954P, 2015MNRAS.453.3649P, 2016MNRAS.456..853P, 2016ApJ...824..100M, 2017A&A...602A..93K}. The resulting $T_\mathrm{eff}$, as shown in the lower panel of Figure~12, drops very quickly after the first two observations from the {\it Swift} UVOT and evolves toward a plateau temperature of $\sim 5300$~K around MJD 57060 ($\sim 30$ days after explosion).

The temporal evolution of $T_\mathrm{eff}$ and $L_\mathrm{bol}$ provide important clues to the properties of SN ejecta and the underlying power source, which helps distinguish various theoretical explanations. In the following analysis, we examine a few possible theoretical models by comparing them with the observed $L_\mathrm{bol}$ and $T_\mathrm{eff}$ of ASASSN-14ms by performing the Markov Chain Monte Carlo (MCMC) algorithms.

\subsection{Radioactive Decay Model (RD)}
 A classic theoretical model is presented by \citet{1982ApJ...253..785A}, where the radioactive decay of $^{56}$Ni powers the ejecta in a homologous expansion with a uniform density distribution. In fitting the observations, we consider the radioactive decay of $^{56}$Ni via $^{56}$Co to $^{56}$Fe as the only energy source (e.g., \citealp{2008MNRAS.383.1485V}) and employ the temperature calculation method described by \citet{2017ApJ...850...55N}. The gamma-ray leakage from the $^{56}$Ni decay is also considered in our model, which modifies the luminosity input by a factor of  $1-\exp(-3\kappa_{\gamma}M_\mathrm{ej}/4\pi v_\mathrm{ej}^2t^2)$ (e.g., \citealp{2015ApJ...799..107W}).

 The free parameters for the model fit include the explosion time ($T_0$), the $^{56}$Ni mass ($M_\mathrm{Ni}$), and the ejecta mass and velocity ($M_\mathrm{ej}$, $v_\mathrm{ej}$). We set $\kappa_{\gamma}=0.027$~cm$^2$~g$^{-1}$ as the opacity for the gamma-ray photons from $^{56}$Ni and $^{56}$Co decay (e.g., \citealp{1997A&A...328..203C, 2000ApJ...545..407M, 2003ApJ...593..931M, 2016ApJ...817..132D}). Based on the MCMC fitting, the best-fit parameters are shown in Table~\ref{Tab: Fit_models}. This model cannot fit the data well (Fig.~12), with $\chi^2/\mathrm{d.o.f}=15$. Moreover, it is physically irrational to have newly-synthesized $^{56}$Ni heavier than the ejecta themselves. Thus, ASASSN-14ms cannot be explained by the radioactive decay model alone, and an additional energy input is needed. A similar conclusion was also drawn by \citet{2018MNRAS.475.2344V} by applying the radioactive-decay model fit to the bolometric light curves of ASASSN-14ms.

 Actually, for a core-collapse SN there is an expected upper limit of 0.2 to the mass ratio between the $^{56}$Ni and ejecta \citep{2008ApJ...673.1014U}. Thus, we take this as one of the criteria for selecting the optimum parameters in the following fitting process with models involving radioactive decay.

\subsection{Magnetar Plus Radioactive Decay Model (MAG+RD)}

A supernova explosion is likely to result in a millisecond magnetar, whose rotation energy can be as high as $\sim10^{52}$ erg. Thus, the spindown of a nascent magnetar is widely accepted as an alternative power source for luminous SNe (e.g., \citealp{2010ApJ...717..245K, 2010ApJ...719L.204W, 2013ApJ...770..128I, 2014MNRAS.444.2096N, 2015ApJ...799..107W, 2016ApJ...821...22W}).

We adopt the magnetar model developed by \citet{2017ApJ...850...55N}, where the recession of the photosphere is taken into account. In addition to the free parameters related to radioactive decay input, this model introduces the dipole magnetic field strength ($B$) and the initial spin period ($P_0$) of a magnetar. In the process of fitting the model, we fix $M_\mathrm{Ni} = 0$ owing to the insignificant contribution of $^{56}$Ni decay to a good fit. The best-fit parameters are reported in Table~6. In this scenario, a rapid rotation in the magnetar can provide a strong wind to power the high luminosity of ASASSN-14ms. Our model requires a magnetar with an initial spin period of 5.6~ms, which is longer than $P_0=1$ ms derived by \citet{2018MNRAS.475.2344V} based on their data, though both values are within the breakup limit of a neutron star (i.e., $\sim0.8$~ms; \citealp{1995A&A...296..745H}). The magnetic field strength of the magnetar is inferred to be $2.87\times10^{14}$ G, which is lower than $B\approx10^{15}$ G obtained by \citet{2018MNRAS.475.2344V}. As shown in Figure~\ref{fig: Compare_LT}, the magnetar model can reproduce the post-peak observations except for a slight deviation at t$\sim$20-40 days after V-band maximum light, with a reduced chi-square $\chi^2/\mathrm{d.o.f.}=3.5$. Moreover, the magnetar model seems to predict a slow rise time for the light curve, which is inconsistent with the fast-rise trend seen in some SNe Ibn. Therefore, the spindown of a newly-born magnetar might not be the main power source for the emission of ASASSN-14ms.

\subsection{Cooling of Shocked Extended Matter}

Inspired by the discoveries of double-peaked light curves in some core-collapse SNe, some studies suggest that cooling emission of shocked extended material may contribute to the first peak preceding the primary one \citep{2014ApJ...788..193N, 2015ApJ...808L..51P, 2020arXiv200708543P}. In this scenario, the extended envelope, with a mass of $M_\mathrm{e}$ and a radius of $R_\mathrm{e}$, is shocked by a blast wave (with an energy of $E_\mathrm{e}$), and then it radiates by cooling. In addition, part of the core (with a mass of $M_\mathrm{c}$) is ejected by the explosion, which sweeps up the extended envelope and they expand outward together. Thus, the total ejecta mass can be defined as $M_\mathrm{ej}=M_\mathrm{c} + M_\mathrm{e}$ (e.g., \citealp{2017ApJ...835...58V}).

Here we fit the observations of ASASSN-14ms with a hybrid model involving shock cooling (SC) \citep{2015ApJ...808L..51P} and radioactive decay (RD) as considered in \S 4.1. The best-fit results are shown in Figure~\ref{fig: Compare_LT} and the corresponding parameters are presented in Table~\ref{Tab: Fit_models} ($\chi^2/\mathrm{dof}=6.9$). Although this model reproduces the late-time luminosity and temperature, it fails to attain a luminosity as high as $10^{44}$ erg s$^{-1}$ at early times. Moreover, this fit requires an unusually large mass ($M_\mathrm{e}=3.32~M_\odot$) and radius ($R_\mathrm{e}=9.8\times10^{13}$ cm) for the envelope of the progenitor star, which is inconsistent with the properties of WR stars as implied by other SNe~Ibc. It is worth noting that the inferred radius is close to the upper limit that we assumed in the fitting.

Another hybrid model (SC+MG), involving shock cooling and spindown of a magnetar, provides a better fit to the observations of ASASSN-14ms ($\chi^2/\mathrm{d.o.f.}=2.6$). In this fit, the early-time emission is primarily due to shock cooling, while a magnetar with $P \approx 5$~ms and $B \approx 6 \times 10^{14}$~G is invoked to explain the late-time light curves. The inferred mass of the stellar envelope ($M_\mathrm{e} = 0.9~M_\odot$) is more reasonable than that required by the SC+RD model. However, similar to the SC+RD model, this model requires an unusually large radius ($R_\mathrm{e} = 9 \times 10^{13}$~cm) for the envelope of the progenitor star.

\subsection{CSM-Ejecta Interaction + Radioactive Decay Model (CSM+RD)}
An exploding star could possibly be surrounded by dense CSM formed through stellar winds, giant eruptions, binary interaction, or a pulsational pair instability (\citealp{2014ARA&A..52..487S, 2017ApJ...836..244W}, and references therein). The interaction between the CSM and outward ejecta would produce a large amount of radiation as well as narrow emission lines as seen in SNe~Ibn, SNe~IIn, and some SLSNe with spectra similar to those of SNe~IIn (for reviews, see \citealp{2017bookG, 2019ARA&A..57..305G}).

In this scenario, we explore the possibility that ASASSN-14ms was powered by not only radioactive decay (see \S\ref{subsec: rd}), but also the forward and reverse shocks produced by the interaction of ejecta with a steady-state wind ($\rho\propto r^{-2}$; e.g., \citealp{2012ApJ...746..121C, 2013ApJ...773...76C, 2018MNRAS.475.2344V, 2019MNRAS.489.1110W}). The steady-state wind could be characterized by mass $M_\mathrm{CSM}$, inner radius $r_1$, and density at inner radius $\rho_\mathrm{CSM,1}$. Here we consider ejecta divided at a dimensionless radius of $x_0=0.7$, and adopt $\delta=2$ and $n=7$, which are power-law exponents of the inner and outer density profiles, respectively. In addition, $\epsilon$ is introduced to represent the efficiency of converting the kinetic energy into radiation during the interaction process.

As one can see, the CSM+RD model provides a better fit for the evolution of $L_\mathrm{bol}$ and $T_\mathrm{eff}$ (Fig.~\ref{fig: Compare_LT}) in comparison with other models, with a relatively small $\chi^2/\mathrm{d.o.f.} = 2.2$; the best-fit model parameters are listed in Table~\ref{Tab: Fit_models}. We note that this model is also favored for ASASSN-14ms according to the fit by \citet{2018MNRAS.475.2344V}. However, they focused on the fits to the bolometric light curve while our model consider fits to both bolometric luminosity and effective temperature, which results in different choice about the best-fit parameters. For example, they derive the masses of the ejecta and CSM as $M_\mathrm{ej}=4.28M_\odot$ and $M_\mathrm{CSM}=0.51M_\odot$, respectively, which are almost half of those inferred from our model ($M_\mathrm{ej}\approx9.0M_\odot$ and $M_\mathrm{CSM}=0.9M_\odot$). In addition, they require an obviously larger $^{56}$Ni ($M_\mathrm{Ni}=0.23M_\odot$) than our result ($M_\mathrm{Ni}=0.04M_\odot$). Our fitting results based on the CSM+RD model suggest that ASASSN-14ms originated from a progenitor star with a high mass-loss rate, $6.7(v_\mathrm{w}/1000$~km~s$^{-1})~M_\odot$~yr$^{-1}$. In principle, it is difficult for a single red supergiant star to drive such a powerful wind except that it suffered intense pulsational ejections before exploding (i.e., \citealp{2010ApJ...717L..62Y}). Although binary interaction could be responsible for such a strong outflow, the velocity expected for this outflow (i.e., 10--100~km~s$^{-1}$) is quite slow compared to that of the He\,\textsc{i} lines of ASASSN-14ms ($\sim1000$~km~s$^{-1}$; see \S~\ref{subsec: spec}). Another competitive progenitor candidate is a luminous blue variable (LBV), which could generate substantial mass loss during its eruptive phases, with a mass-loss rate of $\lesssim 10~M_\odot$ s$^{-1}$ (\citealp{2006ApJ...644.1151S}). The typical velocity of the wind driven by WR/WN stars is $v_\mathrm{w}\sim$100--1000~km~s$^{-1}$ (\citealp{2019MNRAS.488.3783B}, and references therein), in accordance with the He\,\textsc{i} velocity observed in ASASSN-14ms.

Among the well-studied sample of SNe~Ibn, SN 2006jc was observed to undergo a giant LBV-like eruption two years before the SN explosion, consistent with a WR progenitor star evolved from an LBV \citep{2007ApJ...657L.105F, 2007Natur.447..829P}. Note that the outbursts detected shortly before some SNe~IIn resemble the activity of LBVs, which favors the origin of some SNe~IIn from LBVs (e.g., \citealp{2009Natur.458..865G, 2011ApJ...732...32F, 2011ApJ...732...63S, 2017A&A...599A.129T}). A recent study of SN 2017hcc (SN~IIn) indicates that it may have had an LBV progenitor with a mass-loss rate of $0.12 (v_\mathrm{w}/20$~km~s$^{-1})~M_\odot$~yr$^{-1}$ before the explosion \citep{2019MNRAS.488.3089K}. SN 2011hw may represent a transitional event between an SN~IIn and an SN~Ibn, characterizing by narrow emission features of both helium and hydrogen in the spectra. The presence of some residual H in the spectra of SN 2011hw suggests that its precursor was likely a post-LBV star --- a WN-type star \citep{2012MNRAS.426.1905S, 2015MNRAS.449.1954P}. The similarity between SN 2006jc and SN 2011hw suggests a possibility that their progenitors were both in the transition stage from an LBV to a WR star but exploded at different phases \citep{2012MNRAS.426.1905S}.

Therefore, we propose that ASASSN-14ms likely originated from a WR star transited from an LBV, and the surrounding CSM was ejected during an eruption of a WN-type star before the explosion. This scenario could be true, as there have already been detections of LBV-like eruptions from WR stars, such as MCA-1B observed in M33 (\citealp{2020MNRAS.492.5897S}, and references therein). A WR+LBV binary is an alternative progenitor system: ejecta driven by the explosion of a WR star expand and finally catch up with CSM ejected by an LBV (e.g., \citealp{2007Natur.447..829P}), which might lead to a bright transient analogous to ASASSN-14ms. The MCMC fitting results favor this hybrid scenario involving two power sources, suggesting that ASASSN-14ms could result from an explosion of a post-LBV WN-type star (i.e., type WN9-11 or Opfe). It is worth noting that there is inevitably model parameter degeneracy that is hard to break owing to so many parameters involved in this model. Thus, the best-fit results presented here are representative but possibly not the only set of parameters that can reproduce the observations.

\section{Conclusions}
We present extensive optical observations of the Type Ibn supernova ASASSN-14ms, covering the phases from $\sim 2$ days before to $\sim 130$ days after maximum light. Among the known sample of SNe~Ibn, ASASSN-14ms is likely the most luminous event, especially in the UV bands, with $M_{U} < -22.0$ at its peak. Its post-peak light curves exhibit rapid decline, similar to the behavior seen in most SNe~Ibn. However, the narrow emission lines of He~I seem to persist in the spectra for a longer time than in normal SNe~Ibn, suggesting that the CSM distribution is more extended around ASASSN-14ms.

Based on several popular theoretical models proposed for the emission of SNe~Ibn, we attempt to fit the evolution of the bolometric luminosity and effective temperature. According to the fitting results, we find that ASASSN-14ms cannot be primarily powered by radioactive decay, spindown of a nascent magnetar, or cooling of a shocked stellar envelope. On the other hand, the combination of CSM-ejecta interaction and radioactive decay can well account for the observations. Based on the inferred properties for the CSM, we speculate that the progenitor of ASASSN-14ms might have experienced a giant eruption with a high mass-loss rate of $6.7(v_\mathrm{w}/1000$~km~s$^{-1})~M_\odot$~yr$^{-1}$. Such an intense mass-loss rate and wind velocity cannot be simultaneously explained by the stellar wind of a red supergiant or WR star or by the binary interaction. Instead, the helium-rich material around ASASSN-14ms was likely ejected during the late WR (WN9-11 or Opfe) stage.

We also examine the light-curve properties for a sample of well-observed SNe~Ibn and find that their peak luminosities are roughly correlated with the post-peak decline rates, with more luminous SNe~Ibn tending to have slower magnitude decline rates as measured within 30 days after maximum brightness. Considering the role of CSM interaction in affecting the light curves, this correlation may reflect differences in the surrounding CSM (i.e., extension and density distribution) as a result of the distinct evolution of massive stars.

{\acknowledgments} We are grateful to the staffs of various telescopes and observatories with which the data were obtained (Tsinghua-NAOC 0.8-m Telescope, Lijiang 2.4-m Telescope, Xinglong 2.16~m Telescope, Keck 10-m telescope). We thank S. Benetti for providing the archival spectra of SN 1991D. Financial support for this work has been provided by the National Science Foundation of China (NSFC grants 12033003 and 11633002), the Major State Basic Research Development Program (grant 2016YFA0400803), and the Scholar Program of Beijing Academy of Science and Technology (DZ:BS202002). This work was partially supported by the Open Project Program of the Key Laboratory of Optical Astronomy, National Astronomical Observatories, Chinese Academy of Sciences. A.V.F. is grateful for financial assistance from the TABASGO Foundation, the Christopher R. Redlich Fund, and the U.C. Berkeley Miller Institute for Basic Research in Science (where he is a Miller Senior Fellow). K.M. acknowledges support from the Japan Society for the Promotion of Science (JSPS) KAKENHI grant JP18H05223, JP20H00174, and JP20H04737. Y.-Z. Cai is funded by China Postdoctoral Science Foundation (grant no. 2021M691821). Funding for the LJT has been provided by the Chinese Academy of Sciences and the People's Government of Yunnan Province. The FOCAS observation, S15A-078, was conducted at the Subaru telescope, and the authors thank Miho Kawabata, Kentaro Aoki, and Takashi Hattori for obtaining it. We also thank the BASS (Beijing-Arizona Sky Survey) for sharing time on the Bok telescope. Some of the data presented herein were obtained at the W. M. Keck Observatory, which is operated as a scientific partnership among the California Institute of Technology, the University of California, and NASA; the observatory was made possible by the generous financial support of the W. M. Keck Foundation.

{}


\begin{figure}[htbp]
\hspace{4.0cm}
\begin{center}
\includegraphics[angle=0,width=0.8\textwidth]{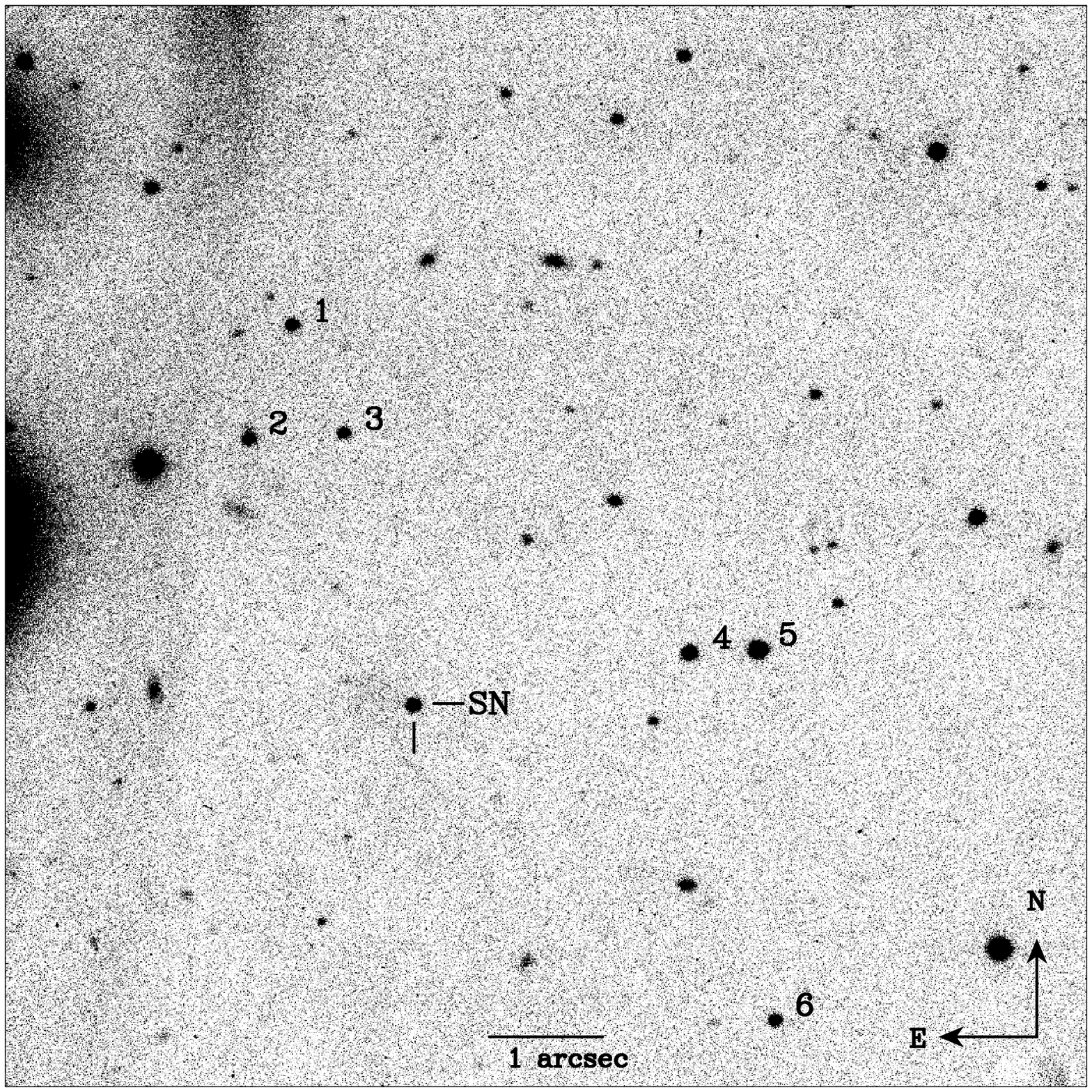}
\end{center}
\vspace{0.2cm}
\caption{$V$-band image of the field of ASASSN-14ms taken with the 0.8~m Tsinghua-NAOC telescope (TNT).
The local standard stars used for flux calibration are marked.}
\label{fig-1}
\vspace{-0.0cm}
\end{figure}

\begin{figure}[htbp]
\hspace{4.0cm}
\begin{center}
\includegraphics[angle=0,width=1.0\textwidth]{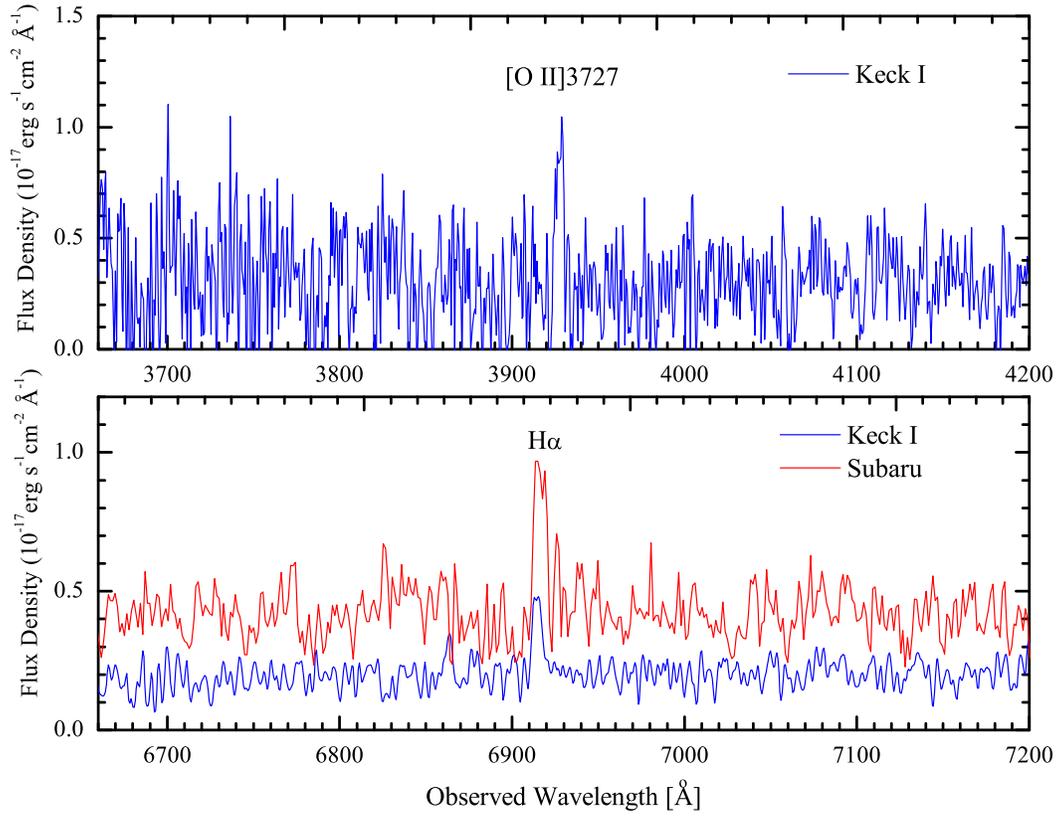}

\end{center}
\vspace{0.2cm}
\caption{The spectra of SDSS J130408.52+521846.4 taken with the Keck-I and Subaru telescopes. The upper panel shows the wavelength region centered on the [O\,\textsc{ii}] $\lambda$3727 line, and the lower panel shows the portion centered on the H$\alpha$ emission.}
\label{fig-2}
\vspace{-0.0cm}
\end{figure}

\begin{figure*}[htbp]
\center
\includegraphics[angle=0,width=1.0\textwidth]{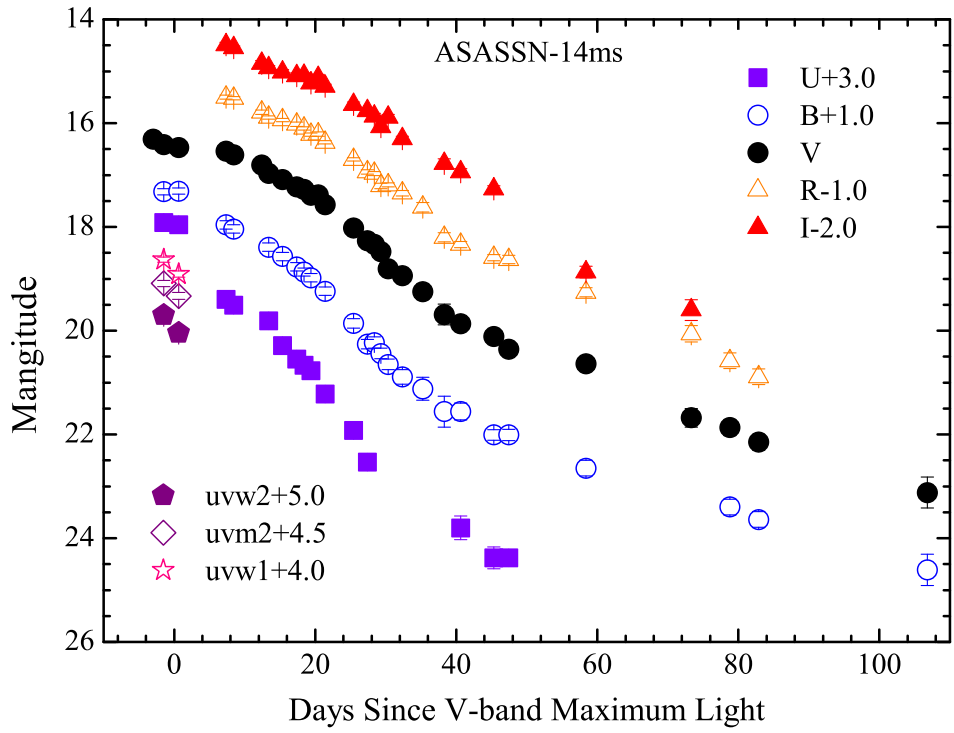} 
\vspace{0.2cm}
\caption{Light curves of ASASSN-14ms obtained with the 0.8~m TNT, 2.4~m LJT, 2.3~m Bok, and {\it Swift} UVOT.}
\label{fig-3}
\vspace{-0.0cm} 
\end{figure*}

\begin{figure}[htbp]
\center
\includegraphics[angle=0,width=1\textwidth]{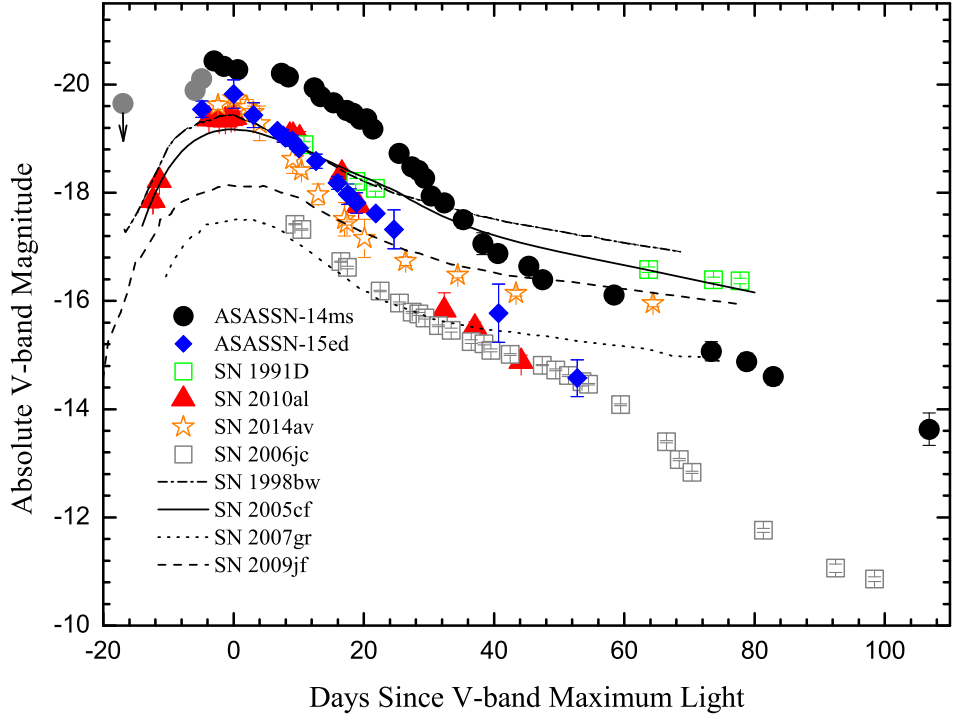} 
\vspace{0.2cm}
\caption{Comparison of the $V$-band light curves of ASASSN-14ms with other SNe~Ibn and representative SNe of different types. Two early-time detections and one prediscovery upper limit from the ASAS-SN (Vallely et al. 2018) are included (grey dots). See text for references.}
\label{fig-4}
\vspace{-0.0cm} 
\end{figure}

\clearpage
\begin{figure}[htbp]
\center \includegraphics[angle=0,width=0.8\textwidth]{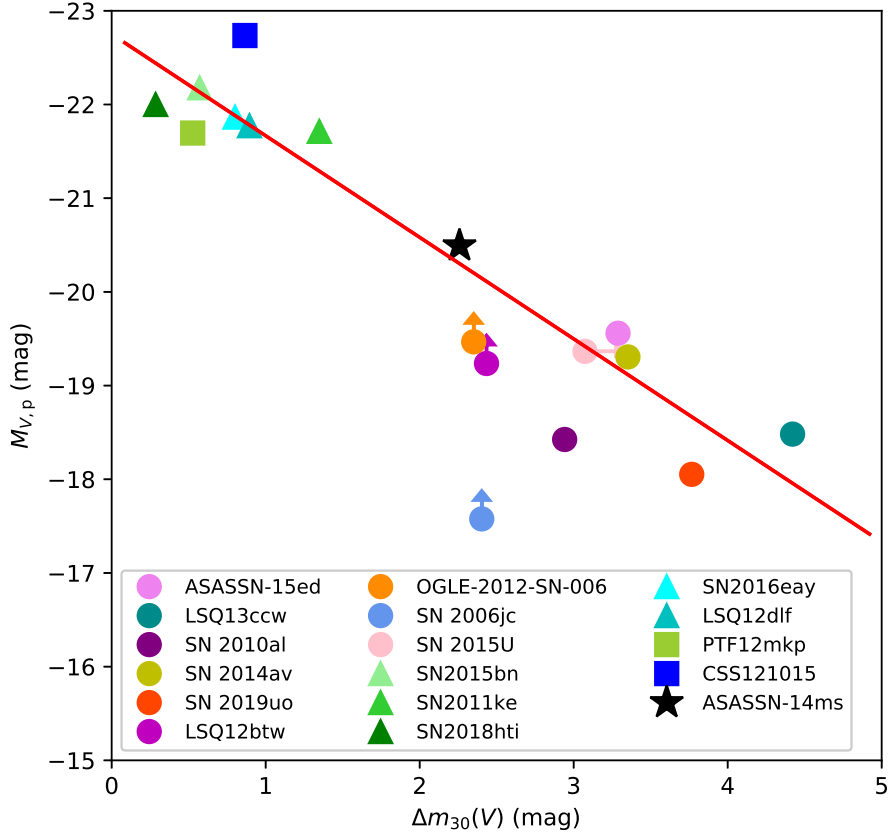}
\vspace{-1.0cm}
\caption{Peak magnitude versus decline rate of $V$-band light curves of ASASSN-14ms (star) as well as some other representative SNe~Ibn (circles), SLSNe~I (triangles), and SLSNe~II (squares). The red line represents the best-fit linear correlation  $M_{V,\mathrm{p}}=-22.75^{+0.06}_{-0.02}+1.08^{+0.01}_{-0.02}\times\Delta m_{30}$. The data are corrected for Galactic extinction. For SN 2015U, high host-galaxy reddening has been discovered in several independent studies \citep{2015MNRAS.454.4293P, 2016MNRAS.461.3057S, 2017ApJ...836..158H}; the value $E(B-V)=0.9^{+0.1}_{-0.4}$ mag with $R_V=2.1$ \citep{2016MNRAS.461.3057S} is used in this paper. Data references: SNe~Ibn: ASASSN-15ed \citep{2015MNRAS.453.3649P}, LSQ12btw \citep{2015A&A...579A..40S}, LSQ13ccw \citep{2015A&A...579A..40S}, SN 2010al \citep{2015MNRAS.449.1921P}, SN 2014av \citep{2016MNRAS.456..853P}, SN 2015U \citep{2016MNRAS.461.3057S}, SN 2006jc \citep{2007Natur.447..829P}, SN 2019uo \citep{2020ApJ...889..170G}, OGLE-2012-SN-006 \citep{2015MNRAS.449.1941P}; SLSNe~I: LSQ12dlf \citep{2015A&A...579A..40S}, SN 2011ke \citep{2013ApJ...770..128I}, SN 2015bn \citep{2016ApJ...826...39N}, SN 2016eay \citep{2014Ap&SS.354...89B}, SN 2018hti \citep{2020MNRAS.497..318L}; SLSNe~II: PTF12mkp \citep{2014Ap&SS.354...89B}, CSS121015-004244+132827 \citep{2015A&A...579A..40S}.}
\label{fig-5} \vspace{-0.0cm} 
\end{figure}

\clearpage
\begin{figure}[htbp]
\center
\includegraphics[angle=0,width=0.7\textwidth]{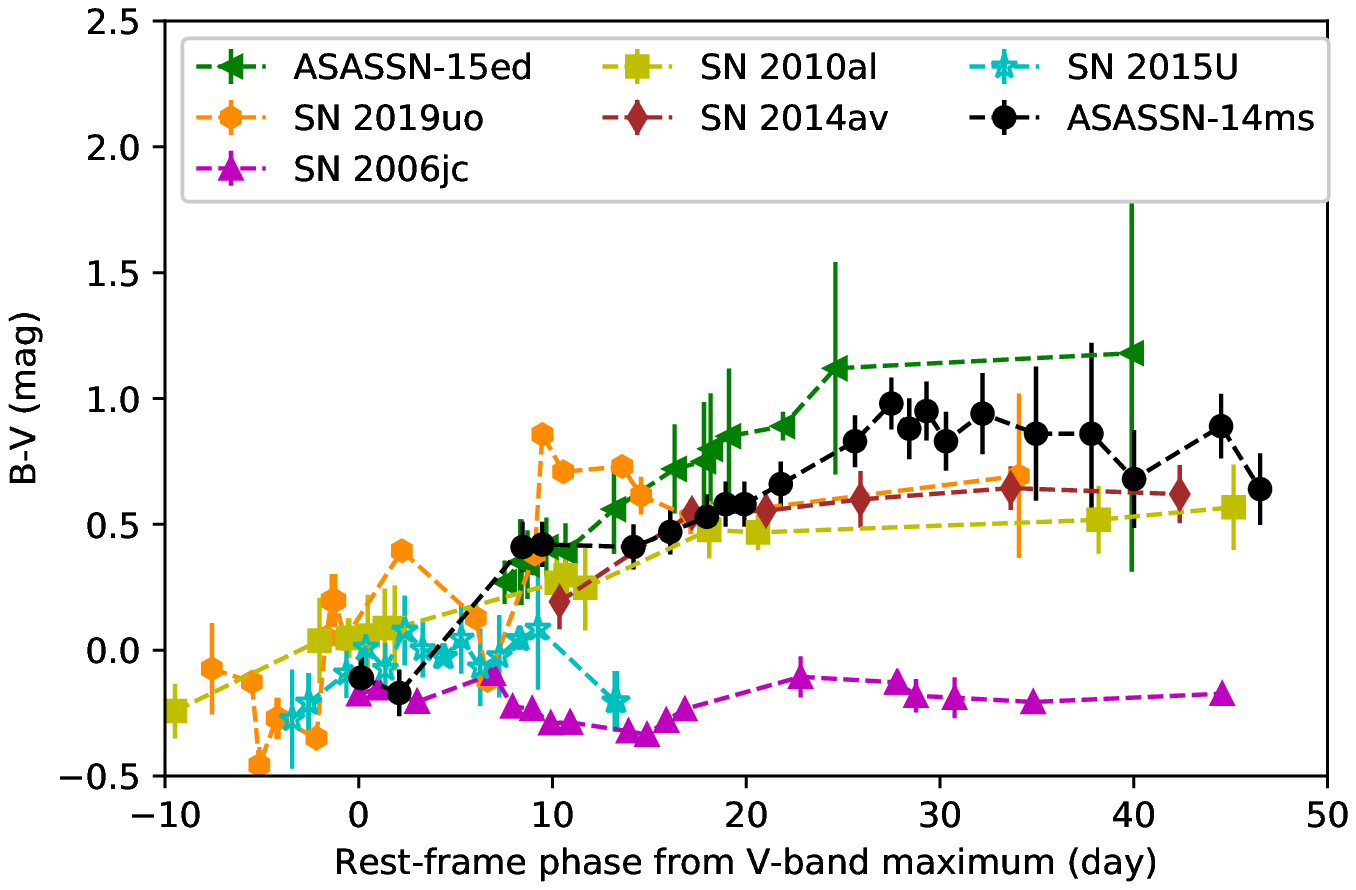}
\includegraphics[angle=0,width=0.7\textwidth]{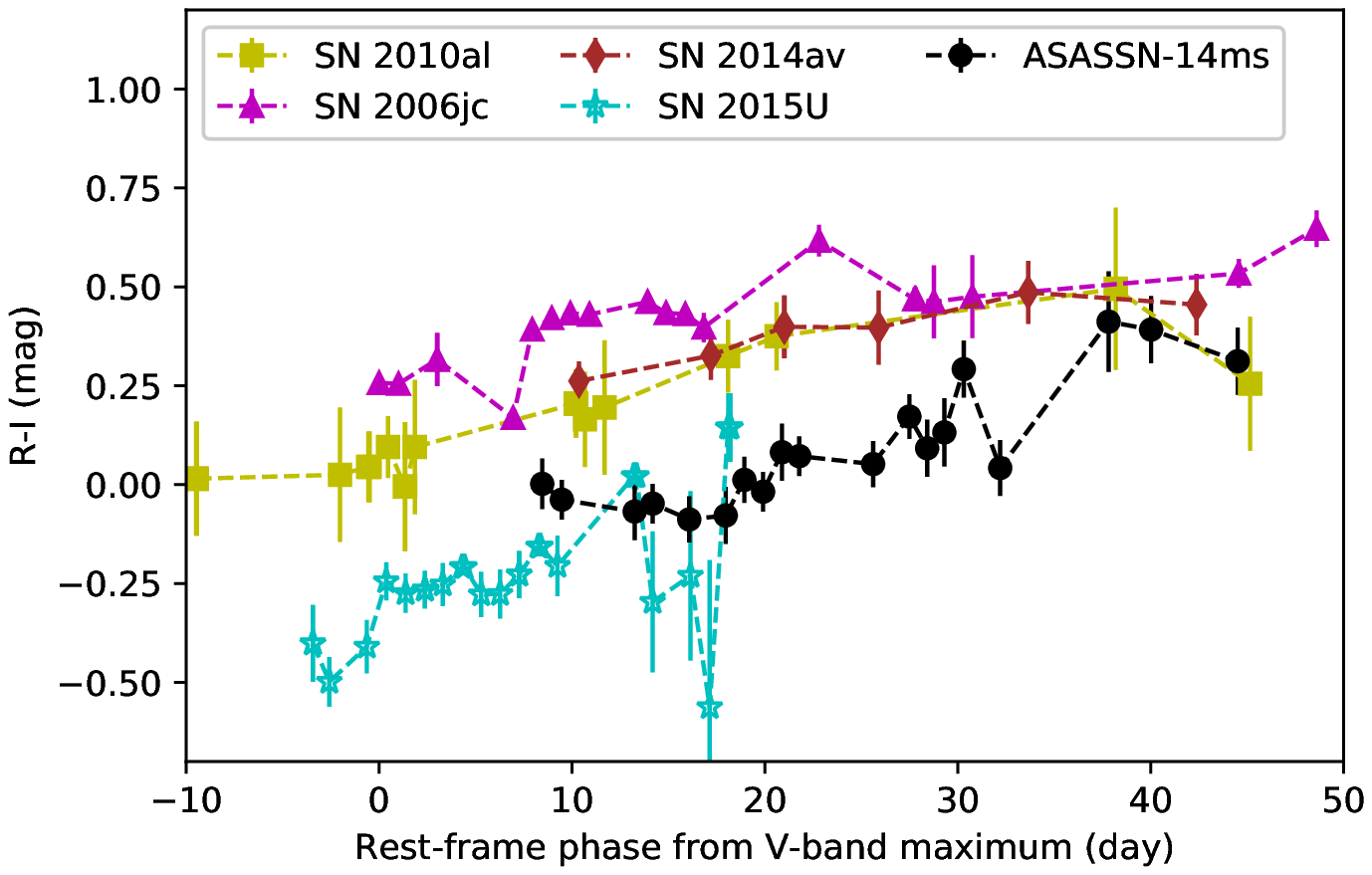}
\vspace{-0.0cm}
\caption{$B-V$ (upper) and $R-I$ (bottom) color of ASASSN-14ms and some well-observed SNe~Ibn. The color curves are corrected for Galactic extinction. The photometry of SN 2015U is corrected for its significant host-galaxy reddening \citep{2016MNRAS.461.3057S}. Data references: ASASSN-15ed \citep{2015MNRAS.453.3649P}, SN2006jc \citep{2007Natur.447..829P}, SN 2010al \citep{2015MNRAS.449.1921P}, SN 2014av \citep{2016MNRAS.456..853P}, SN 2015U \citep{2016MNRAS.461.3057S}, SN2019uo \citep{2020ApJ...889..170G}.}
\label{fig-6}
\end{figure}

\clearpage 
\begin{figure}[htbp]
\center
\includegraphics[angle=0,width=1\textwidth]{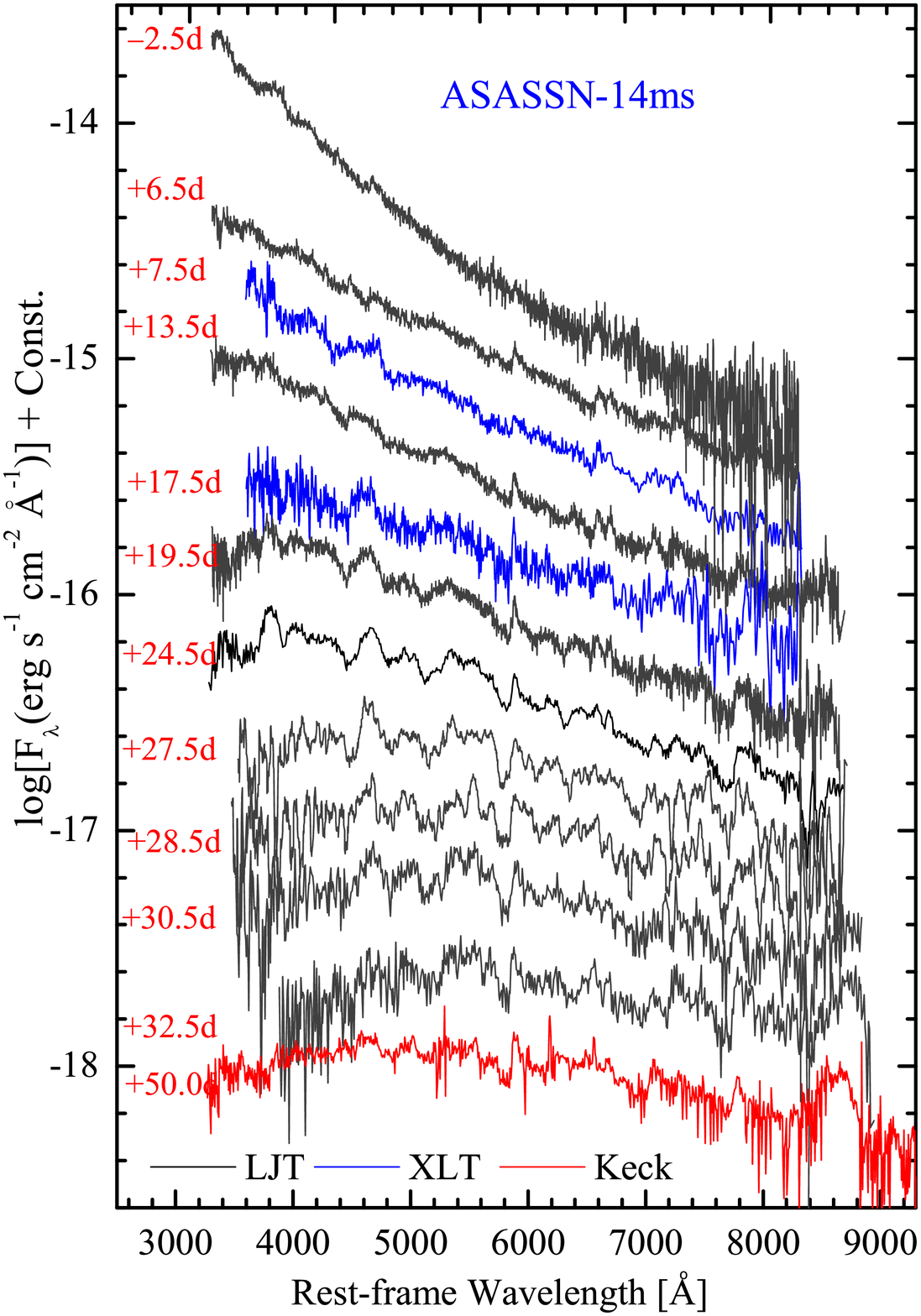}
\vspace{-2.5cm} 
\caption{Optical spectra of ASASSN-14ms. Spectra obtained with the YNAO 2.4~m telescope are shown in black, those obtained with the Xinglong 2.16~m of NAOC are in blue, and the Keck spectrum is shown in red.}
\label{fig-7}
\end{figure}

\clearpage
\begin{figure}[htbp]
\center
\includegraphics[angle=0,width=0.9\textwidth]{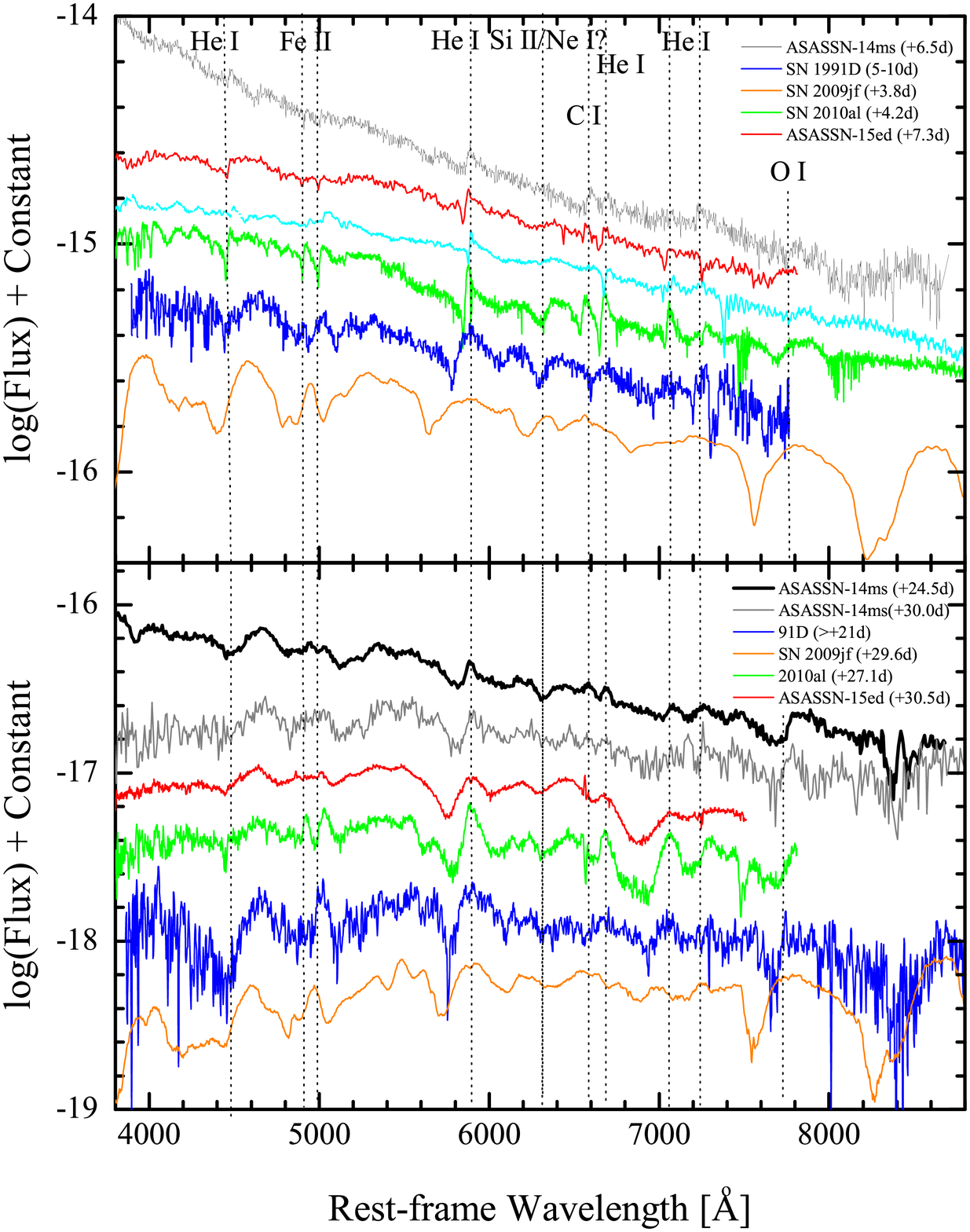}
\vspace{-2.0cm}
\caption{Spectral comparison between ASASSN-14ms and some well-observed SNe~Ibn and SNe~Ib, including ASASSN-15ed, SN 2010al, SN 1991D, and SN 2009jf. The dotted lines mark the spectral features identified in those objects, including He\,\textsc{i} lines, Fe\,\textsc{ii}, O\,\textsc{i} $\lambda$7774, and possibly Si\,\textsc{ii}/Ne\,\textsc{i}.}
\label{fig-8}
\vspace{-0.0cm}
\end{figure}

\clearpage
\begin{figure}[htbp] \center
\includegraphics[angle=0,width=0.8\textwidth]{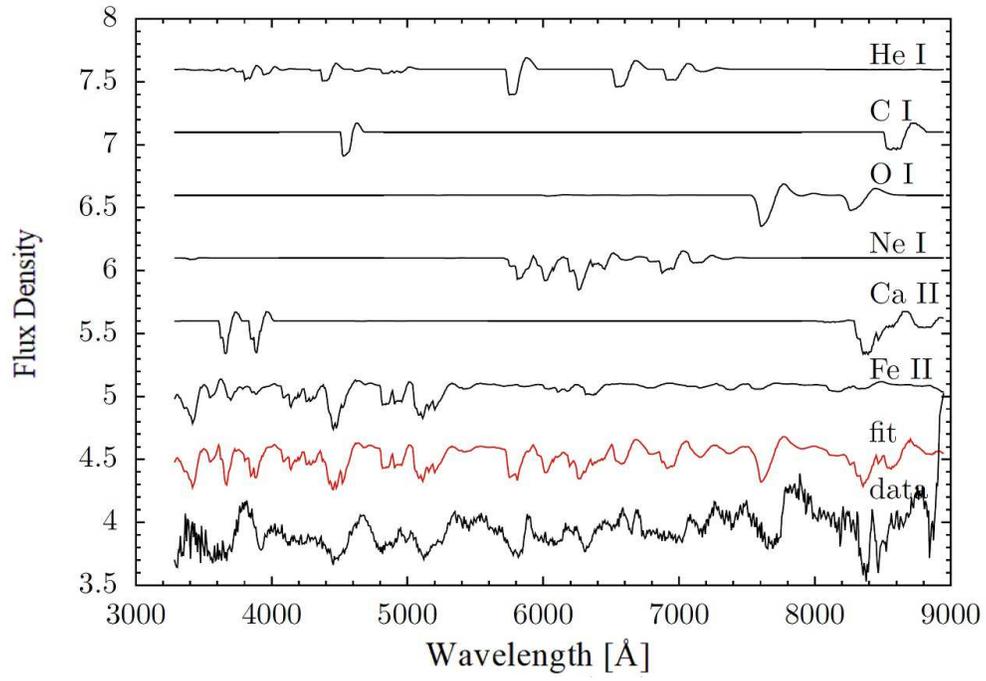} 
\vspace{0.2cm}
\caption{The SYNAPPS fit to the t$\sim$+24.5 day spectrum of ASASSN-14ms.}
\label{fig-9}
\vspace{-0.0cm} 
\end{figure}

\clearpage 
\begin{figure}[htbp]
\center \includegraphics[angle=0,width=1\textwidth]{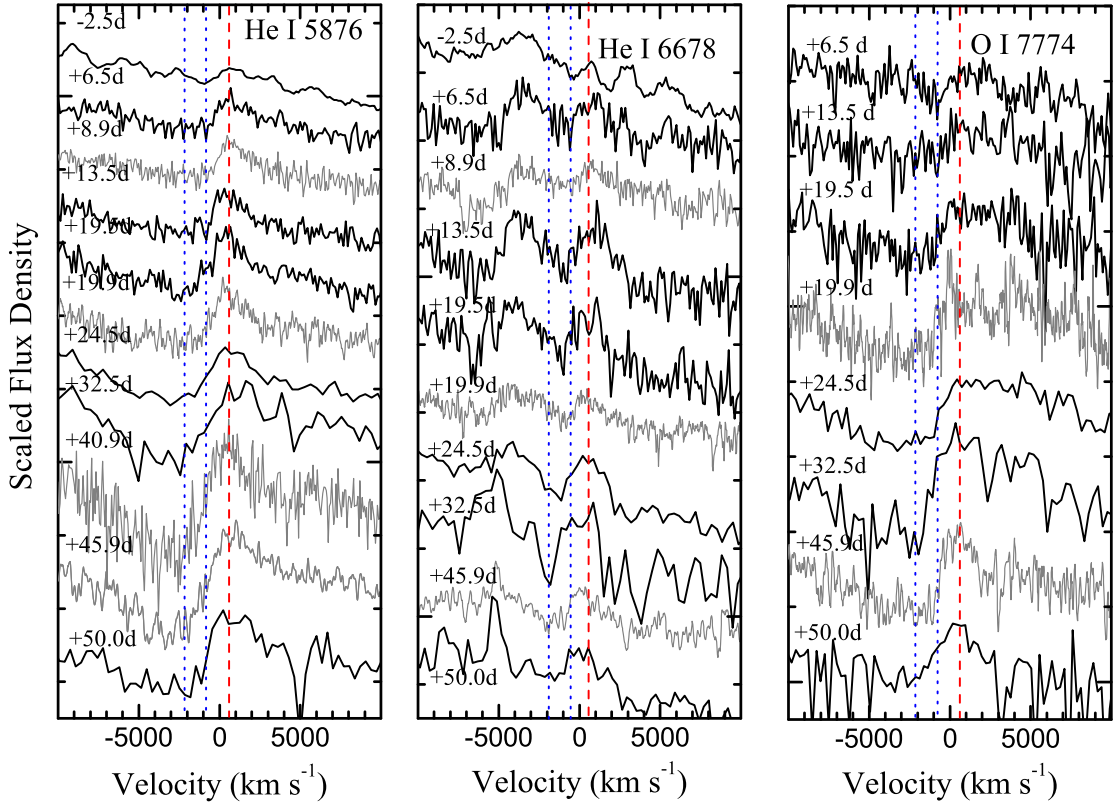}
\vspace{-0.0cm}
\caption{Temporal evolution of line profiles of He\,\textsc{i} $\lambda$5876, He\,\textsc{i} $\lambda$6678, and O\,\textsc{i} $\lambda$7774 seen in the optical spectra of ASASSN-14ms; those spectra shown with gray colors at several epochs (i.e., at t$\sim$+8.9d, +19.9d, +40.9d, +45.9d) are taken from Vallely et al. (2018). Note that our determination of the $V-$band maximum-light time is +1.9 days earlier than that derived by Vallely et al. (2018). Red dash-dotted lines mark the rest position of the emission, and the blue lines represent the position of absorption minima at velocities of $\sim 800$ km s$^{-1}$ and 2000 km s$^{-1}$.}
\label{fig-10} \vspace{-0.0cm} 
\end{figure}

\clearpage
\begin{figure}[htbp] \center
\includegraphics[angle=0,width=0.8\textwidth]{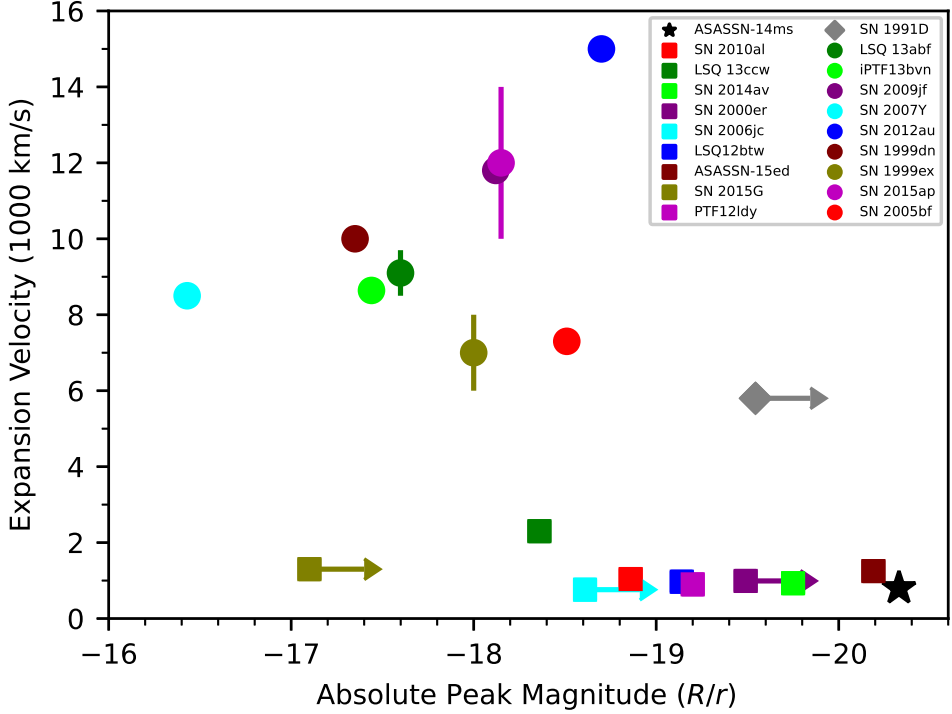} 
\vspace{0.2cm}
\caption{Expansion velocities plotted as a function of the $R/r$-band absolute magnitudes at maximum brightness for ASASSN-14ms and other SNe~Ibn and SNe~Ib. Note that we adopted the $V$-band maximum as an approximation of the $R$ band for ASASSN-14ms, as its R-band light curve was not well sampled around the peak and the $V-R$ color is roughly zero in the early phases. Lower limits are marked by symbols with arrows. Expansion velocities were measured from the spectra at an epoch similar to the $R/r$-band maxima ($0 \pm 15$ days). We preferentially measured the position of the deep absorption produced by He\,\textsc{i} $\lambda$5876, and assuming negligible contribution from Na\,\textsc{i}~D $\lambda\lambda$5890, 5896. In a few SNe~Ibn (P~Cygni profile weak or absent), the values were inferred from the FWHM of the narrow component of the strongest He\,\textsc{i} lines \citep{2017ApJ...836..158H}. The data for the comparison sample are from \citet{2002AJ....124..417H,2002AJ....124.2100S, 2002MNRAS.336...91B, 2003fthp.conf..222H, 2005ApJ...631L.125A, 2009ApJ...696..713S, 2007PASP..119..135P, 2011MNRAS.413.2583S, 2011MNRAS.416.3138V,2011MNRAS.411.2726B, 2013ApJ...772L..17T, 2014MNRAS.445.1932S, 2016MNRAS.456..853P,2019MNRAS.485.1559P, 2020A&A...634A..21S,2020MNRAS.tmp.2067G}}.
\label{fig-11}
\vspace{-0.0cm} 
\end{figure}

\clearpage 
\begin{figure}[htbp] \center
\includegraphics[angle=0,width=0.7\textwidth]{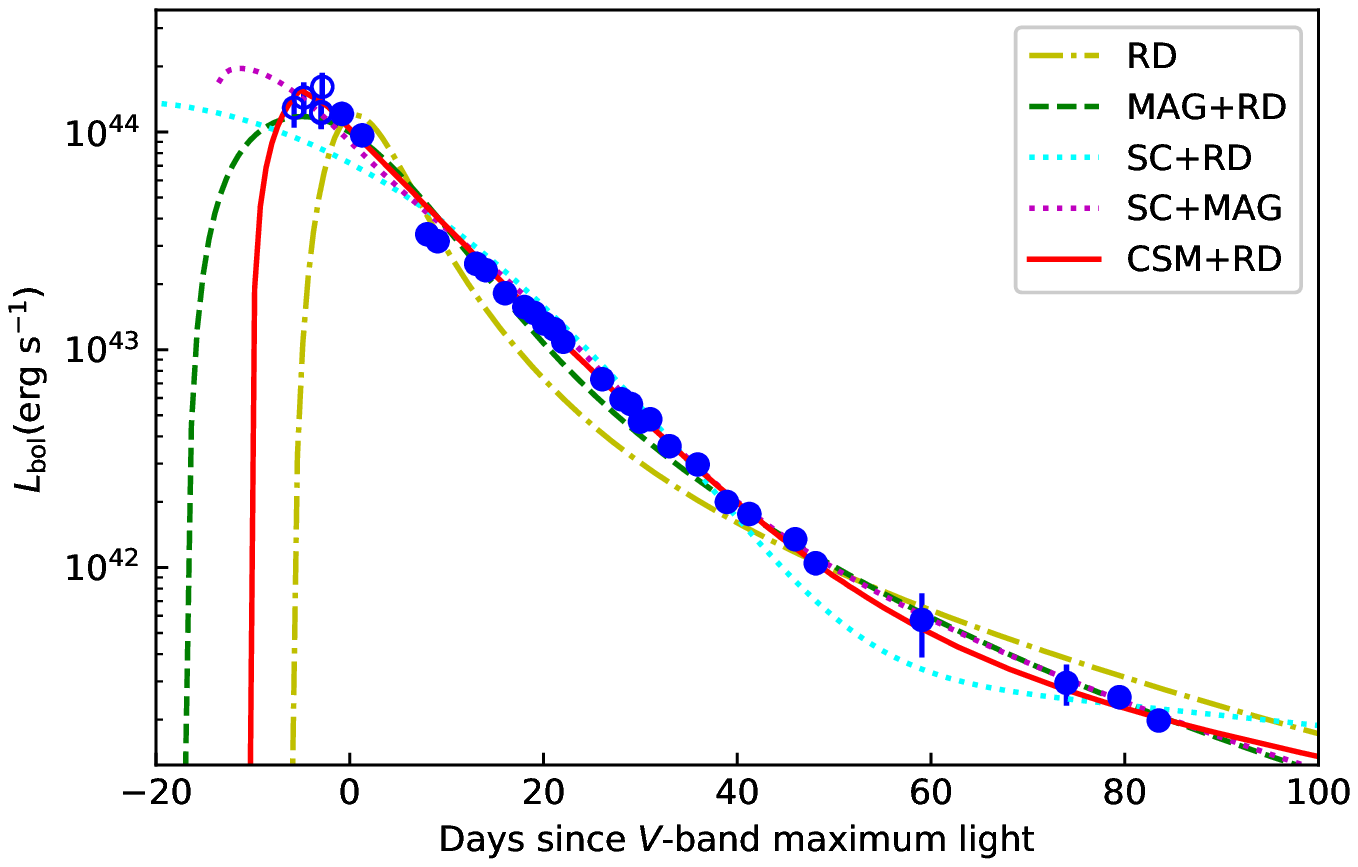} 
\includegraphics[angle=0,width=0.7\textwidth]{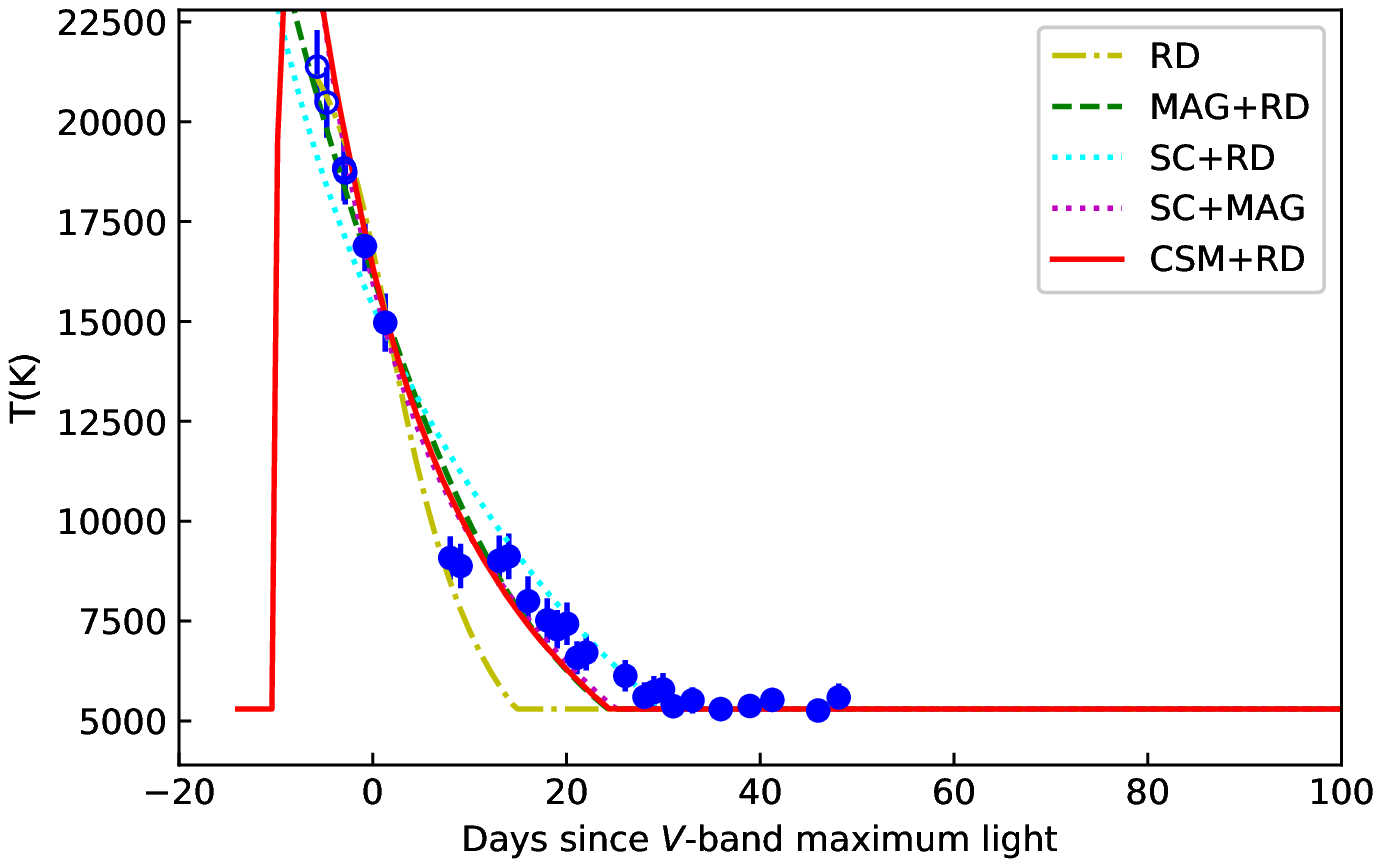}
\vspace{0.2cm}
\caption{Bolometric luminosity (upper) and effective temperature (lower) of ASASSN-14ms. Filled dots represent the values derived from multiband photometry, while open circles represent those inferred from single-band photometry with an assumption of linear decline for temperature. We also show the fitting results of four theoretical models: RD (yellow), MAG (green), SC+RD (cyan), SC+MAG (purple), and CSM+RD (red).}
\label{fig-12}
\vspace{-0.0cm} 
\end{figure}

\clearpage 
\begin{deluxetable}{cccccccc}
\hspace{-1.0cm}
\tablecolumns{6}
\tablewidth{0pc}
\tabletypesize{\scriptsize}
\tablecaption{Photometric Standards in the ASASSN-14ms Field\tablenotemark{a}}
\tablehead{\colhead{Star} & \colhead{$\alpha$(J2000)} & \colhead{$\delta$(J2000)} & \colhead{$U$ (mag)} & \colhead{$B$ (mag)} & \colhead{$V$ (mag)} & \colhead{$R$ (mag)} & \colhead{$I$ (mag)} }
\startdata
1& 13:04:15.13 & +52:22:04.80& 20.12(08)   & 19.00(02)  & 17.73(02)  & 16.75(02)  &16.06(02) \\
2& 13:04:17.70 & +52:21:05.56& 19.87(07)   & 18.81(02)  & 17.57(02)  & 16.83(02)  &16.28(02) \\
3& 13:04:12.32 & +52:21:07.93& 20.31(09)   & 19.19(02)  & 17.91(02)  & 17.28(02)  &16.69(02) \\
4& 13:03:52.89 & +52:19:10.39& 17.84(03)   & 17.62(01)  & 16.90(01)  & 16.45(02)  &15.97(02) \\
5& 13:03:48.98 & +52:19:11.57& 17.78(03)   & 16.81(01)  & 15.63(01)  & 14.96(01)  &14.97(01) \\
6& 13:03:48.38 & +52:15:58.40& 18.04(03)   & 18.23(02)  & 17.76(02)  & 17.57(02)  &17.22(03)
\enddata
\tablenotetext{a}{See Figure~1 for a chart of ASASSN-14ms and the comparison stars. Uncertainties, in units of 0.01 mag, are $1\sigma$.}
\end{deluxetable}

\begin{deluxetable}{cccccccc}
\hspace{-1.0cm}
\tablecolumns{8}
\tablewidth{0pc}
\tabletypesize{\scriptsize}
\tablecaption{Optical Photometry of ASASSN-14ms\tablenotemark{a}}
\tablehead{\colhead{UT Date} & \colhead{JD$-$2,450,000} & \colhead{$U$ (mag)} &\colhead{$B$ (mag)} &\colhead{$V$ (mag)} &\colhead{$R$ (mag)} &\colhead{$I$ (mag)}  & \colhead{Telescope}}
\startdata

20141215 & 7007.03 &     ...       &     ...       &   $>$17.10   &     ...      &     ...        &    ASAS-SN  \\
20141226 & 7018.11 &     ...       &     ...       &   16.86(14)\tablenotemark{b}   &     ...      &     ...        &   ASAS-SN  \\
20141227 & 7019.10 &     ...       &     ...       &   16.64(11)\tablenotemark{b}   &     ...      &     ...        &   ASAS-SN  \\
20141229 & 7021.02	&	  ...       &	  ...   	&   16.31(07)	&	  ...    	 &	  ...  	   &  Stan Howerton	 \\
20141230 & 7022.54 &   14.92(05)   &   16.32(06)   &   16.42(07)   &     ...       &     ...       &    LJT   \\
20150101 & 7024.64 &   14.96(05)   &   16.31(06)   &   16.47(07)   &     ...       &     ...       &    LJT   \\
20150108 & 7031.35	&	16.40(11)   &	16.96(08)   &	16.54(06)	&	16.50(04)	&	16.49(05)	&	 TNT	\\
20150109 & 7032.42	&	16.51(10)   &	17.04(08)   &	16.61(04)	&	 16.52(03)	&	16.55(04)	&	 TNT	\\
20150113 & 7036.40	&	  ...       &	  ...  	    &	16.81(08)	&	16.79(04)	&	 16.85(06)	&	 TNT	\\
20150114 & 7037.38	&	16.81(10)	&	17.39(08)	&	16.97(04)	&	16.89(03)	&	16.93(04)	&	 LJT \\
20150116 & 7039.38	&	17.29(10)	&	17.57(08)	&	17.09(04)	&	16.94(03)	&	17.02(05)	&	 TNT \\
20150118 & 7041.37	&	 17.55(10)	&	17.77(08)	&	17.23(04)	&	17.02(04)	&	17.09(06)	&	 TNT \\
20150119 & 7042.39	&	17.67(10)	&	17.87(08)	 &	17.28(04)	&	17.10(03)	&	17.08(05)	&	 TNT \\
20150120 & 7043.41	&	17.77(11)	&	17.98(08)	&	17.39(04)	&	17.22(03)	&	17.23(04)	&	 LJT \\
20150121 & 7044.44	&	  ...  	    &	  ...  	    &	17.38(06)	&	17.21(04)	&	17.12(06)  &	 TNT \\
20150122 & 7045.39	&	18.22(10)	&	18.24(08)	&	17.57(04)	&	17.37(03)	&	17.29(04)	&	 LJT	\\
20150126 & 7049.42	&	18.92(12)	&	18.86(09)	&	18.02(05)	&	17.70(03)	&	17.64(05)	&	 LJT \\
20150128 & 7051.40	&	19.53(16)	&	 19.26(09)	&	18.27(05)	&	17.94(04)	&	17.76(04)	&	 LJT \\
20150129 & 7052.37	&	  ...  	    &	19.23(11)	&	18.34(05)	&	17.97(04)	&	17.87(06)	&	 LJT \\
20150130 & 7053.31	&	  ...  	    &	19.44(10)	&	18.48(06)	&	18.21(05)	&	 18.07(07)	&	 TNT \\
20150131 & 7054.37	&	  ...  	    &	19.65(10)	&	18.81(06)	&	18.19(04)	&	17.89(06)	&	 TNT	\\
20150202 & 7056.36	&	  ...    	&	19.89(14)	&	18.94(08)	&	18.35(05)	&	18.30(05)	&	 LJT	\\
20150205 & 7059.27	&	  ...    	&	20.12(22)	&	19.25(15)	&	18.62(09)	&	  ...    	&	 TNT	\\
20150208 & 7062.28	&	  ...   	&	20.56(30)	 &	19.69(20)	&	19.20(09)	&	18.78(09)	&	 LJT    \\
20150211 & 7064.60	&	20.80(23)	&	20.56(16)	&	19.87(11)	&	19.34(06)	&	18.94(06)	&	 LJT	\\
20150215 & 7069.33	&	21.38(21)	&	21.01(10)	&	20.11(08)	&	19.59(06)	&	 19.27(06)	&	 LJT	\\
20150217 & 7071.45	&	21.38(14)	&	21.01(11)	&	20.36(09)	&	19.63(08)	&	  ...    	&	 LJT	 \\
20150228 & 7082.41	&	  ...   	&	21.65(15)	&	20.64(11)	&	20.26(09)	&	20.87(11)	&	 LJT	\\
20150315 & 7097.34	 &	  ...   	&	  ...   	&	21.68(18)	&	21.06(16)	&	21.60(20)	&	 LJT	\\
20150321 & 7102.80	&	  ...   	&	 22.40(15)	&	21.87(15)	&	21.58(15)	&	  ...   	&	 BoK	\\
20150325 & 7106.86	&	  ...   	&	22.64(15)	&	 22.15(15)	&	21.89(15)	&	  ...   	&	 BoK	\\
20150418 & 7130.80	&	  ...   	&	23.61(30)	&	23.12(30)	&	  ...   	 &	  ...   	&	 BoK	\\
20150515 & 7157.60	&	  ...       &	24.40(100)	&	23.70(100)	&	  ...   	&	  ...   	&	 BoK	\\

\enddata
\tablenotetext{a}{The epoch of $V$-band maximum is 7023.85.
  Uncertainties, in units of 0.01 mag, are $1\sigma$.}
\tablenotetext{b}{Data are taken from \citet{2018MNRAS.475.2344V}.}
\end{deluxetable}

\begin{deluxetable}{ccccccccc}
\hspace{-1.0cm}
\tablecolumns{8}
\tablewidth{0pc}
\tabletypesize{\scriptsize}
\tablecaption{{\it Swift} UVOT Photometry of ASASSN-14ms\tablenotemark{a}}
\tablehead{\colhead{Date} & \colhead{JD$-$2,450,000} &\colhead{$uvw2$ (mag)} & \colhead{$uvm2$ (mag)} & \colhead{$uvw1$ (mag)} & \colhead{$u$ (mag)} &\colhead{$b$ (mag)} &\colhead{$v$ (mag)}}
\startdata
20141230 & 7022.04	&	14.70(05)   &	 14.59(06)   &   14.63(05) &	14.92(05)   &   16.32(06)   &   16.42(07)  \\
20150101 & 7024.14	&  15.04(06)   &	 14.83(07)   &   14.91(06) &    14.96(05)   &   16.31(06)   &   16.47(07)  \\
\enddata
\tablenotetext{a}{Uncertainties, in units of 0.01 mag, are $1\sigma$.}
\label{tab-Swift}
\end{deluxetable}

\begin{deluxetable}{cccccc}
\hspace{-1.0cm}
\tablecolumns{6}
\tablewidth{0pc}
\tabletypesize{\scriptsize}
\tablecaption{Journal of Spectroscopic Observations of ASASSN-14ms}
\tablehead{\colhead{UT Date} & \colhead{JD$-$2,450,000} & \colhead{Phase\tablenotemark{a}} & \colhead{Exp. (s)} & \colhead{Telescope + Instrument} & \colhead{Range (\AA)}}
\startdata
2014 Dec. 29 &	7021.36 &   $-$2.49 &	1500 &	YNAO 2.4 m+YFOSC(G3) &	3500--8750 \\
2015 Jan. 07 &	7030.39 &   +6.54  &	1800 &	YNAO 2.4 m+YFOSC(G3) &	3500--8750 \\	
2015 Jan. 08 &	7031.31 &   +7.46  &	2400 &	BAO 2.16 m+BFOSC(G4) &	3800--8800 \\	
2015 Jan. 14 &	7037.40 &   +13.55 &	1500 &	YNAO 2.4 m+YFOSC(G3) &	3500--8750 \\
2015 Jan. 18 &	7041.32 &   +17.47 &	3600 &	BAO 2.16 m+BFOSC(G4) &	3800--8800 \\
2015 Jan. 20 &	7043.41 &   +19.56 &	2100 &	YNAO 2.4 m+YFOSC(G3) &	3500--8750 \\
2015 Jan. 25 &	7048.39 &   +24.54 &	3000 &	YNAO 2.4 m+YFOSC(G10) &	3700--9500 \\
2015 Jan. 29 &	7052.33 &   +28.48 &	2400 &	YNAO 2.4 m+YFOSC(G3) &	3500--8750 \\
2015 Jan. 31 &	7054.43 &   +30.58 &	2400 &	YNAO 2.4 m+YFOSC(G10) &	4000--9300 \\
2015 Feb. 02 &	7056.37 &   +32.52 &	2100 &	YNAO 2.4 m+YFOSC(G10) &	4000--9600 \\
2015 Feb. 22 &  7076.93 &   +53.08 &    1200 &  Keck LRIS             & 3000--10,000  \\
2015 Jun. 21 &  7194.90 &   +171.05 &   1800 &  Subaru FOCAS          & 3500--9500  \\

\enddata
\tablenotetext{a}{Relative to the epoch of $V$-band maximum (JD = 2,457,023.85).}
\end{deluxetable}

\begin{deluxetable}{lcccccc}
\hspace{-1.0cm}
\tablecolumns{2}
\tablewidth{0pc}
\tabletypesize{\scriptsize}
\tablecaption{Input SYNAPPS fit parameters for t$\sim$+24.5 day spectrum of ASASSN-14ms}
\tablehead{\colhead{Parameters} & \colhead{He I} &\colhead{C I} &\colhead{O I} &\colhead{Ne I} &\colhead{Ca II} &\colhead{Fe II}}
\startdata
 log$_{\tau}$(line opacity)      & $-$2.1 & $-$0.7& $-$2.5 & $-$2 & $-$5& $-$6 \\
 v$_{min}$(10$^{3}$ km s$^{-1}$)       &  0.1 & 0.1 & 1 & 1 & 0.1 & 1 \\
 v$_{max}$(10$^{3}$ km s$^{-1}$)        &  7.9 & 7.5 & 8.0 & 10 & 7.5 & 7.5 \\
 v$_{e}$(10$^{3}$ km s$^{-1}$)         &  0.4 & 0.8 & 0.4 & 0.6 & 0.1 & 0.6  \\
 temp(10$^{3}$ K)    &9 & 1 & 8 & 8 & 3 & 1 \\
\enddata
\end{deluxetable}

\begin{deluxetable}{ccccccccccc}
\tablecolumns{10}
\tablewidth{0pc}
\hspace{-1.0cm}
\tabletypesize{\scriptsize}
\tablecaption{Best-fit model parameters.}
\setlength{\tabcolsep}{1.5mm}
\tablehead{\colhead{Model} & \colhead{RD} & \colhead{MAG}& \colhead{SC+RD}& \colhead{SC+MAG} &\colhead{CSM+RD}  }
\startdata
$T_0$ (MJD)                   &  57017.3        & 57006.3   & 57009.9   & 57009.7       & 57009.3          \\
$\kappa$ (cm$^2$~g$^{-1}$)  & 0.19  &0.18 & 0.17 & 0.19  &0.16\\
$M_\mathrm{ej}~(M_\odot)$     &  0.43            & 1.13   &-- &--       & 8.99             \\
$v_\mathrm{ej}$ (km~s$^{-1}$) &  $2.76\times10^4$  & $9.7\times10^3$ & $1.08\times10^4$ &
$1.18\times10^4$ & $1.17\times10^4$  \\
$M_\mathrm{Ni}~(M_\odot)$     &  3.9  & 0     &0.05 &--        & 0.04            \\
$B$ ($10^{14}$ G)             &  --                & 2.87 &--  & 6.73  &--\\
$P_0$ (ms)                    &  --                & 5.6 &--  &  2.5 &--\\
$\kappa_{\gamma,m}$\tablenotemark{a} (cm$^2$~g$^{-1}$) & -- & 0.01 &-- & 0.038 &--\\
$M_\mathrm{c}~(M_\odot)$  & -- & -- &  3.56 & 1.66 & --\\
$M_\mathrm{e}~(M_\odot)$  & -- & --  & 3.32 & 0.9 & --\\
$R_\mathrm{e}$ ($10^{13}$ cm)  & -- & --  & 9.7 & 9  & --\\
$E_\mathrm{SN}$ ($10^{51}$ erg)  & -- & --   & 1.77 & 1.6 & --\\
$M_\mathrm{CSM}~(M_\odot)$    & -- & -- & -- & -- & 0.9  \\
$\rho_\mathrm{CSM,1}$ (g~cm$^{-3}$) & -- & -- & -- & -- & $2.22\times10^{-12}$
 \\
$r_1$ (cm)    & -- & --  & -- & -- & $ 3.9\times10^{14}$   \\
$\epsilon$    & -- & --  & -- & -- &   0.75   \\ \hline
$\chi^2/$d.o.f  & 15 & 3.5  & 6.9 & 2.6 &  2.2   \\
\enddata
\tablenotetext{a}{The opacity for $\gamma$-ray photons from a magnetar.}
\end{deluxetable}

\end{document}